\documentclass[usenatbib]{mnras}

\usepackage{enumitem}
\usepackage{newtxtext,newtxmath}
\usepackage[T1]{fontenc}
\usepackage{ae,aecompl}
\usepackage{amsmath}
\usepackage{graphicx}
\usepackage[utf8]{inputenc}
\usepackage{qtree}
\usepackage[normalem]{ulem}
\usepackage{gensymb}
\usepackage[table]{xcolor}

\newcommand{\abar}{\mbox{$a_\mathrm{bar}$}}

\newcommand{\bn}{\mbox{$b_\mathrm{n}$}}

\newcommand{\cs}{\mbox{$c_\mathrm{s}$}}
\newcommand{\cssqr}{\mbox{$c_\mathrm{s}^2$}}

\newcommand{\dMBH}{\mbox{$\dot{M}_\mathrm{BH}$}}
\newcommand{\dMBHL}{\mbox{$\dot{M}_\mathrm{BHL}$}}
\newcommand{\dMBHprime}{\mbox{$\dot{M}_\mathrm{BH}'$}}
\newcommand{\dMradio}{\mbox{$\dot{M}_\mathrm{radio}$}}
\newcommand{\dMEdd}{\mbox{$\dot{M}_\mathrm{Edd.}$}}
\newcommand{\dSFRnorm}{\mbox{$(\delta \mathrm{SFR})_{\mathrm{norm.}}$}}

\newcommand{\epsilonf}{\mbox{$\epsilon_\mathrm{f}$}}
\newcommand{\epsilonr}{\mbox{$\epsilon_\mathrm{r}$}}

\newcommand{\Ie}{\mbox{$I_\mathrm{e}$}}

\newcommand{\kB}{\mbox{$k_\mathrm{B}$}}

\newcommand{\Lradio}{\mbox{$L_\mathrm{radio}$}}
\newcommand{\LX}{\mbox{$L_\mathrm{X}$}}

\newcommand{\MBH}{\mbox{$M_\mathrm{BH}$}}
\newcommand{\MBHprime}{\mbox{$M_\mathrm{BH}'$}}
\newcommand{\MBHsqr}{\mbox{$M_\mathrm{BH}^2$}}

\newcommand{\Md}{\mbox{$M_\mathrm{d}$}}

\newcommand{\mH}{\mbox{$m_\mathrm{H}$}}

\newcommand{\mpr}{\mbox{$m_\mathrm{pr}$}}
\newcommand{\MSersic}{\mbox{$M_\mathrm{S\acute{e}rsic}$}}

\newcommand{\nelectron}{\mbox{$n_\mathrm{e}$}}
\newcommand{\nH}{\mbox{$n_\mathrm{H}$}}

\newcommand{\Omegab}{\mbox{$\Omega_{\mathrm{b}}$}}
\newcommand{\Omegam}{\mbox{$\Omega_\mathrm{m}$}}

\newcommand{\rBH}{\mbox{$r_\mathrm{BH}$}}
\newcommand{\Reff}{\mbox{$R_\mathrm{eff.}$}}
\newcommand{\re}{\mbox{$r_\mathrm{e}$}}

\newcommand{\SFRfiducial}{\mbox{$\mathrm{SFR}_{\mathrm{fiducial}}$}}
\newcommand{\SFRvariant}{\mbox{$\mathrm{SFR}_{\mathrm{variant}}$}}
\newcommand{\sigmaT}{\mbox{$\sigma_\mathrm{T}$}}

\newcommand{\tlookback}{\mbox{$t_\mathrm{lookback}$}}
\newcommand{\Tvir}{\mbox{$T_\mathrm{vir}$}}

\newcommand{\upsilonBH}{\mbox{$\upsilon_\mathrm{BH}$}}
\newcommand{\upsilonBHsqr}{\mbox{$\upsilon_\mathrm{BH}^2$}}

\newcommand{\aquarius}{\mbox{\textsc{Aquarius}}}
\newcommand{\arepo}{\mbox{\textsc{Arepo}}}
\newcommand{\astropy}{\mbox{\textsc{Astropy}}}
\newcommand{\auriga}{\mbox{\textsc{Auriga}}}
\newcommand{\eagle}{\mbox{\textsc{Eagle}}}
\newcommand{\eris}{\mbox{\textsc{Eris}}}
\newcommand{\erisBH}{\mbox{\textsc{ErisBH}}}
\newcommand{\imfit}{\mbox{\textsc{Imfit}}}
\newcommand{\matplotlib}{\mbox{\textsc{Matplotlib}}}
\newcommand{\numpy}{\mbox{\textsc{NumPy}}}
\newcommand{\photutils}{\mbox{\textsc{Photutils}}}
\newcommand{\scipy}{\mbox{\textsc{SciPy}}}

\newcommand{\Mstar}{\hbox{$M_\mathrm{\star}$}}
\newcommand{\Msun}{\hbox{$\mathrm{M_\odot}$}}

\newenvironment{shortitem}
{\begin{list}{$\bullet$}{\topsep=0pt\itemsep=0pt\parsep=0pt\parskip=0pt\leftmargin=12pt}}
{\end{list}}
\newcommand{\bsi}{\begin{shortitem}}
\newcommand{\esi}{\end{shortitem}}
\newenvironment{shortsubitem}
{\begin{list}{$\circ$}{\topsep=0pt\itemsep=0pt\parsep=0pt\parskip=0pt\leftmargin=12pt}}
{\end{list}}
\newcommand{\bssi}{\begin{shortsubitem}}
\newcommand{\essi}{\end{shortsubitem}}

\newcommand{\App}[1]{Appendix~\ref{app:#1}}
\newcommand{\Eq}[1]{equation~(\ref{eq:#1})}
\newcommand{\Fig}[1]{Fig.~\ref{fig:#1}}
\newcommand{\Sec}[1]{Section~\ref{sec:#1}}
\newcommand{\Tab}[1]{Table~\ref{tab:#1}}

\begin{document}

\title[The effects of AGN feedback on MW-mass galaxies]{The effects of AGN feedback on the structural and dynamical properties \\ of Milky Way-mass galaxies in cosmological simulations}
\author[D. Irodotou et al.]
{{Dimitrios Irodotou,$^{1,2}$}\thanks{E-mail: Dimitrios.Irodotou@helsinki.fi}
Francesca Fragkoudi,$^{3,4,5}$
Ruediger Pakmor,$^5$
Robert J.J. Grand,$^{5,6,7}$
\newauthor
Dimitri A. Gadotti,$^4$
Tiago Costa,$^5$
Volker Springel,$^5$
Facundo A. G\'omez$^{8,9}$ and
Federico Marinacci$^{10}$ \\
$^1$Department of Physics, University of Helsinki, Gustaf H$\ddot{a}$llstr$\ddot{o}$min katu 2, FI-00014, Helsinki, Finland \\
$^2$Astronomy Centre, University of Sussex, Falmer, Brighton BN1 9QH, UK \\
$^3$Institute for Computational Cosmology, Department of Physics, Durham University, Durham DH1 3LE, UK \\
$^4$European Southern Observatory, Karl-Schwarzschild-Str. 2, 85748 Garching, Germany \\
$^5$Max-Planck-Insitut f$\ddot{u}$r Astrophysik, Karl-Schwarzschild-Str. 1, 85748 Garching, Germany \\
$^6$Instituto de Astrof\'isica de Canarias, Calle Vía L\'actea s/n, E-38205 La Laguna, Tenerife, Spain\\
$^7$Departamento de Astrof\'isica, Universidad de La Laguna, Av. del Astrof\'isico Francisco S\'anchez s/n, E-38206, La Laguna, Tenerife, Spain\\
$^8$Instituto de Investigaci\'on Multidisciplinar en Ciencia y Tecnolog\'ia, Universidad de La Serena, Ra\'ul Bitr\'an 1305, La Serena, Chile\\
$^9$Departamento de Astronom\'ia, Universidad de La Serena, Av. Juan Cisternas 1200 Norte, La Serena, Chile\\
$^{10}$Department of Physics \& Astronomy ``Augusto Righi'', University of Bologna, via Gobetti 93/2, 40129 Bologna, Italy\\
}
\pagerange{\pageref{firstpage}--\pageref{lastpage}} \pubyear{2021}
\maketitle

\label{firstpage}
\begin{abstract}
Feedback from active galactic nuclei (AGN) has become established as a fundamental process in the evolution of the most massive galaxies. Its impact on Milky Way (MW)-mass systems, however, remains comparatively unexplored. In this work, we use the \auriga\ simulations to probe the impact of AGN feedback on the dynamical and structural properties of galaxies, focussing on the bar, bulge, and disc. We analyse three galaxies -- two strongly and one unbarred/weakly barred -- using three setups: (i) the fiducial \auriga\ model, which includes both radio and quasar mode feedback, (ii) a setup with no radio mode, and (iii) one with neither the radio nor the quasar mode. When removing the radio mode, gas in the circumgalactic medium cools more efficiently and subsequently settles in an extended disc, with little effect on the inner disc. Contrary to previous studies, we find that although the removal of the quasar mode results in more massive central components, these are in the form of compact discs, rather than spheroidal bulges. Therefore, galaxies without quasar mode feedback are more baryon-dominated and thus prone to forming stronger and shorter bars, which reveals an anti-correlation between the ejective nature of AGN feedback and bar strength. Hence, we report that the effect of AGN feedback (i.e. ejective or preventive) can significantly alter the dynamical properties of MW-like galaxies. Therefore, the observed dynamical and structural properties of MW-mass galaxies can be used as additional constraints for calibrating the efficiency of AGN feedback models.
\end{abstract}


\section{Introduction} \label{sec:Introduction}

Observational evidence suggests that bars exist in the centres of over half of local spiral galaxies \citep{EFP00,AMC09,NA10,BSA15,E18}. From a theoretical perspective, bars can form in isolated galaxies via discs instabilities \citep{H71,OP73,ELN82} or in interacting discs through tidal forces \citep{N96,LAD14}, and subsequently grow in mass and size by redistributing angular momentum \citep{A03} and capturing disc stars \citep{AM02,KGA16}.

Numerous studies have identified significant correlations between galactic and bar properties, which reveal causal connections between the two \citep[e.g.][]{A99,LSR02,KSK12,AMR13}. Moreover, in most late-type barred galaxies, bars coexist with multiple components, such as bulges and super-massive black holes (SMBH) \citep{KK04,K10,A13}, and each one affects and is affected by the others in non-linear ways. For example, while bars evolve they push gas to the centre which can in principle both promote the formation of pseudo-bulges \citep{SFB89,SM99,K13,FDF15,GCD15,FAB16} and fuel the accretion disc around the SMBH \citep{L69,SBF90,SR98,B04,MN12}. As a response, powerful AGN feedback can suppress the gas inflow \citep[e.g.][]{RMN06} and halt the growth of the pseudo-bulge. Hence, bars not only significantly alter the structural properties of galaxies, but also take part in a complex interplay with other components.

Energetic feedback is the natural culprit for shutting off the formation of stars via (at least) two modes, namely the quasar and the radio mode, which are differentiated by their distinct effect. The former is responsible for directly ejecting gas, whereas the latter for preventing gas from cooling effectively \citep[e.g.][]{DB06,F12,GB14,CRS18,CPS20,ZPN20}. This is the reason why AGN feedback has been considered a mechanism responsible for the quenching and formation of some early-type galaxies (ETGs). For example, \cite{DPP16,PMS17} and \cite{FNH19} showed that AGN feedback is important in order to produce massive slow-rotating ETGs \citep{ECK07} since it prevents them from developing young fast-rotating stellar discs. In lower mass haloes ($\sim$ 10$^{12}$ \Msun), AGN feedback has been regarded as ineffectual, and is therefore often not included in simulations \citep[e.g.][]{CAR16,RVC16,ZL16}. However, as we will show below, MW-mass galaxies can also be affected by powerful AGN which plays a fundamental role for their properties \citep[see][and references therein]{SHB21} and regulates the formation of their constituent stellar components and gas reservoirs \citep{HES19,VMB20}.

Recent studies have explored the properties of bars in cosmological simulations \citep[e.g.][]{ANA17,PL19,RBD20,RKP21,RBD21}. Furthermore, a series of works by \cite{BMK16,SBD17} and \cite{ZCD19} have studied in more detail the link between AGN feedback and bars. \cite{BMK16} showed that adding AGN feedback to the \eris\ simulation, promoted the formation of a bar in a MW-mass galaxy, by suppressing the growth of a massive spheroidal component at the centre, whose presence would otherwise stabilise the disc and delay bar formation \citep{CDF90,A13}. However, theoretical studies have also shown that, after the bar has formed, a classical bulge is expected to amplify the transfer of angular momentum from the bar to the bulge which results in the bar losing angular momentum, slowing down, and becoming stronger \citep{A03,SMG12}. Thus, classical bulges can both delay bar formation and fuel bar growth after the bar has been able to form. 

While there has been recent work exploring the bar-AGN feedback connection \citep[e.g.][]{BMK16,SBD17,ZCD19,RBD20,RBD21}, there are still a number of open question related to how AGN feedback, and in particular how different modes of this feedback (e.g. ejective versus preventive) affect the dynamics of disc galaxies. In this work, we therefore study the impact of AGN feedback on the structural and dynamical properties of MW-mass galaxies, by exploring in detail the effects of two different modes of AGN feedback in the \auriga\ simulations, the so-called quasar and radio mode. In particular, we assess to what extent the \auriga\ AGN implementation affects the growth and properties of the disc, bar, and bulge. For that purpose, we selected three galaxies -- two strongly and one very weakly barred -- from the \auriga\ project \citep{GGM17} and re-simulated each one with two different AGN feedback setups whilst keeping their initial conditions and other physical processes identical to the fiducial model. The \auriga\ simulations\footnote{\href{https://wwwmpa.mpa-garching.mpg.de/auriga/}{https://wwwmpa.mpa-garching.mpg.de/auriga/}} provide an ideal testbed for our analysis, since they produce realistic barred galaxies with prominent boxy/peanut and pseudo-bulges \citep{GMG19,BFP20,FGP20,FGP21}.

This paper is organised as follows. In \Sec{The Auriga simulations}, we briefly describe the \auriga\ model and its AGN feedback implementation along with the simulations we use for this study. In \Sec{Present day galactic properties}, we present the $z$ = 0 morphologies and properties of the galaxies, and analyse the radial profile of the bar strength and that of the stellar distribution. Finally, we estimate the relative contribution of the bar, bulge, and disc component to the total stellar mass by performing photometric 2D decompositions. In \Sec{Quasar mode effects} and \Sec{Radio mode effects}, we analyse the effects of, respectively, the quasar and radio modes of AGN feedback on the temporal evolution of the galaxies. Finally, in \Sec{Discussion} we discuss our findings in the context of previous results in the literature and summarise our results in \Sec{Conclusions}.

\section{The \auriga\ simulations} \label{sec:The Auriga simulations}

The \auriga\ project\footnote{\auriga\ is a project of the Virgo consortium for cosmological supercomputer simulations \href{http://virgo.dur.ac.uk/index.html}{http://virgo.dur.ac.uk}.} is a suite of cosmological magneto-hydrodynamical zoom-in simulations of isolated MW-mass haloes \citep{GGM17}. These were selected from the dark-matter-only \eagle\ Ref-L100N1504 cosmological volume \citep{SCB15} and re-simulated at higher resolution with the $N$-body, magneto-hydrodynamics code \arepo\ \citep{S10,PBS11,PSB16,WSP20}. The simulations incorporate a detailed galaxy formation model which includes primordial and metal-line cooling \citep{VGS13}, a hybrid multi-phase star formation model \citep{SH03}, supernova feedback in the form of an effective model for galactic winds and mass and metal return from AGB stars and SNIa \citep{MPS14,GGM17}, a redshift dependent and spatially uniform UV background \citep{FCL09,VGS13}, magnetic fields \citep{PS13,PMS14,PGG17,PGP18}, and a black hole formation and feedback model which we describe in more details below.

In this work, we use nine \auriga\ haloes: three fiducial runs, first presented in \cite{GGM17}, namely Au-06, Au-17, and Au-18, plus two additional model variants for each one which were simulated for the needs of this study. These variants are as follows:
\begin{itemize}
\setlength{\itemindent}{0.5em}
\item the \textbf{NoR} (i.e. no-radio) variants have black hole particles (which grow via gas accretion and mergers) and black hole feedback through the quasar mode, but the radio mode feedback is turned off.
\item the \textbf{NoRNoQ} (i.e. no-radio-no-quasar) variants have their black hole particles, and both quasar mode and radio mode feedback, removed.
\end{itemize} 

For this study we use two galaxies, Au-17 and Au-18, which are strongly barred at $z$ = 0 and have been extensively studied in previous works \citep[e.g.][]{GVZ19,BFO20,FGP20,GKB20,FGP21}. In addition, we analyse Au-06 which can be considered unbarred or weakly barred depending on the threshold used (see \Sec{Present day galactic properties:Morphological properties} for our definition of bar strength), which allows us to draw conclusions about the effects of AGN feedback on unbarred and weakly barred cases, as we discuss in \Sec{Discussion:The fate of unbarred galaxies}.

Throughout this work we use the level-4 resolution \citep[based on the \aquarius\ project nomenclature,][]{SWV08} which corresponds to dark matter and baryonic particles masses of 3 $\times$ 10$^5$ \Msun\ and 5 $\times$ 10$^4$ \Msun, respectively. The gravitational co-moving softening length for stellar and dark matter particles is set to 500 $h^{-1}$ cpc and is set to a fixed value of 250 $h^{-1}$ pc from $z$ = 1 onwards. This is equal to the minimum co-moving softening length allowed for gas cells, while their maximum physical softening length can not exceed 1.85 kpc. For more details on the simulations we refer the reader to \cite{GGM17}. In the following subsection we describe the AGN feedback prescription in the \auriga\ simulations.

\subsection{Black hole and AGN feedback model}\label{sec:The Auriga simulations:Black hole and AGN feedback model}

Black holes come into existence by converting the densest gas cell into a collisionless sink particle in Friends-of-Friends (FoF) groups whose mass exceeds 5 $\times$ 10$^{10}$ $h^{-1}$ \Msun. The initial (seed) mass is set to $10^{5}$ $h^{-1}$ \Msun\ which can increase either by acquiring mass from nearby gas cells or by merging with other black hole particles \citep{SDH05,VGS13}. The former is described by an Eddington-limited Bondi-Hoyle-Lyttleton accretion \citep{HL39,BH44,B52} along with a term to account for the radio mode accretion (see below). Hence, the total black hole accretion rate can be expressed by
\begin{flalign} \label{eq:dMBH}
\dMBH = \mathrm{min} \left [\dMBHL + \dMradio, \dMEdd \right] \;, &&
\end{flalign}
where the term \dMEdd\ represents the Eddington accretion rate which is defined as
\begin{flalign} \label{eq:dMEdd}
\dMEdd\ = \frac{4\pi\ G\ \MBH\ \mpr}{\epsilonr\ \sigmaT\ c} \;, &&
\end{flalign}
where $G$ is the gravitational constant, \MBH\ is the black hole mass, \mpr\ is the proton mass, \epsilonr\ = 0.2 is the black hole radiative efficiency parameter, \sigmaT\ is the Thomson cross section, and $c$ is the speed of light.

Finally, for both modes the energy ejected from the black hole is given by
\begin{flalign} \label{eq:dotE}
\dot{E} = \dMBH\ \epsilonf\ \epsilonr\ c^2 \;, &&
\end{flalign}
where \dMBH\ is given by \Eq{dMBH} and \epsilonf\ = 0.07 is the fraction of radiated energy that thermally couples to the gas. For the quasar mode this energy is injected isotropically as thermal energy into neighbouring gas cells; while for the radio mode \citep{SSD07} it takes the form of bubbles (with size 0.1 times the virial radius) which are stochastically inflated at random locations in the halo\footnote{Following an inverse square distance profile around the black hole.} up to a maximum radius of 0.8 times the virial radius.

\subsubsection{Quasar mode}\label{sec:The Auriga simulations:Black hole and AGN feedback model:Quasar mode}

In \Eq{dMBH}, the term \dMBHL\ represents the Bondi-Hoyle-Lyttleton accretion rate which is defined as
\begin{flalign} \label{eq:dMBHL}
\dMBHL =  \frac{4\pi\ G^2\ \MBHsqr\ \rho}{\left( \cssqr + \upsilonBHsqr \right)^{3/2}} \;, &&
\end{flalign}
where $\rho$ and \cs\ are the density and sound speed, respectively, of the surrounding gas, and \upsilonBH\ is the velocity of the black hole relative to the gas. 

\subsubsection{Radio mode}\label{sec:The Auriga simulations:Black hole and AGN feedback model:Radio mode}

In \Eq{dMBH}, the term \dMradio\ represents the radio mode accretion rate which is defined as
\begin{flalign} \label{eq:dMradio}
\dMradio = \frac{\Lradio}{\epsilonf\ \epsilonr\ c^2} \;. &&
\end{flalign}
This term follows a scheme in which first the X-ray luminosity of the halo (\LX) is converted to a radio mode luminosity (\Lradio), which is then used to define the radio mode accretion rate. This scheme establishes a self-regulated feedback mechanism by following the process described below.

We first measure the X-ray luminosity on a per FoF halo basis by following the observed X-ray luminosity–temperature relation \citep{PCA09}
\begin{flalign} \label{eq:LX}
\LX = \frac{H(z)}{H_0} C \left( \frac{T}{T_0} \right)^{2.7} \;, &&
\end{flalign}
where $H(z)$ and $H_0$ are the Hubble parameter and Hubble constant, respectively, $C$ = 6 $\times$ 10$^{44}$ erg s$^{-1}$, and $T_0$ = 5 Kev. Then we follow \cite{NF00} who developed a theory for spherically symmetric accretion of hot gas onto super-massive black holes at the centres of large haloes. This theory connects the accretion rate back to the state of the gas at large distances away from the black hole based on the formula
\begin{flalign} \label{eq:dMBHprime}
\dMBHprime = \frac{2\pi\ Q(\gamma - 1) \kB\ \Tvir}{\mu \mH\ \Lambda(\Tvir)} G\ \MBHprime \left[ \frac{\rho^2}{\nelectron\ \nH} \right]\mathcal{M}^{3/2} \;,  &&
\end{flalign}
where $Q$ = 2.5, $\gamma$ = 5/3 is the adiabatic index of the gas, \kB\ is the Boltzmann constant, \Tvir\ is the virial temperature, $\mu$\mH\ is the mean mass per gas cell, $\Lambda$(\Tvir) is the cooling function, $\rho$, \nelectron, and \nH\ are the mean gas, electron number, and hydrogen number densities, respectively, $\mathcal{M}$ = 0.0075 is the Mach number of the gas far away from the black hole, and \MBHprime\ is derived from the observed \MBH-$\sigma$ relation
\begin{flalign} \label{eq:MBHprime}
\MBHprime(\sigma) = M_0 \left( \frac{\sigma}{\sigma_0} \right)^4 \;, &&
\end{flalign}
where $\sigma$ is the velocity dispersion of stars in the galactic bulge. Lastly, we assume a radio power associated with this radio mode accretion which can be written as
\begin{flalign} \label{eq:Lradio}
\Lradio = R(\Tvir,z) \LX \;, &&
\end{flalign}
where \LX\ is the X-ray luminosity of the halo introduced in \Eq{LX} and we define the ratio
\begin{flalign} \label{eq:R}
R(\Tvir,z) \equiv  \frac{\epsilonf\ \epsilonr\ c^2\ \dMBHprime}{\LX} \;, &&
\end{flalign}
which tends to increase for larger halos and towards later times.

\section{Properties at \lowercase{$z$} = 0} \label{sec:Present day galactic properties}

In this section, we investigate the effects of the AGN feedback (or the lack of it) on the $z$ = 0 stellar surface densities and bar strengths (\Sec{Present day galactic properties:Morphological properties}) and on the total stellar mass (\Sec{Present day galactic properties:Stellar-halo mass relation}) and its distribution (\Sec{Present day galactic properties:Stellar mass distribution}).

\subsection{Morphological properties} \label{sec:Present day galactic properties:Morphological properties}

\begin{figure*}
\centering \includegraphics[width=\textwidth]{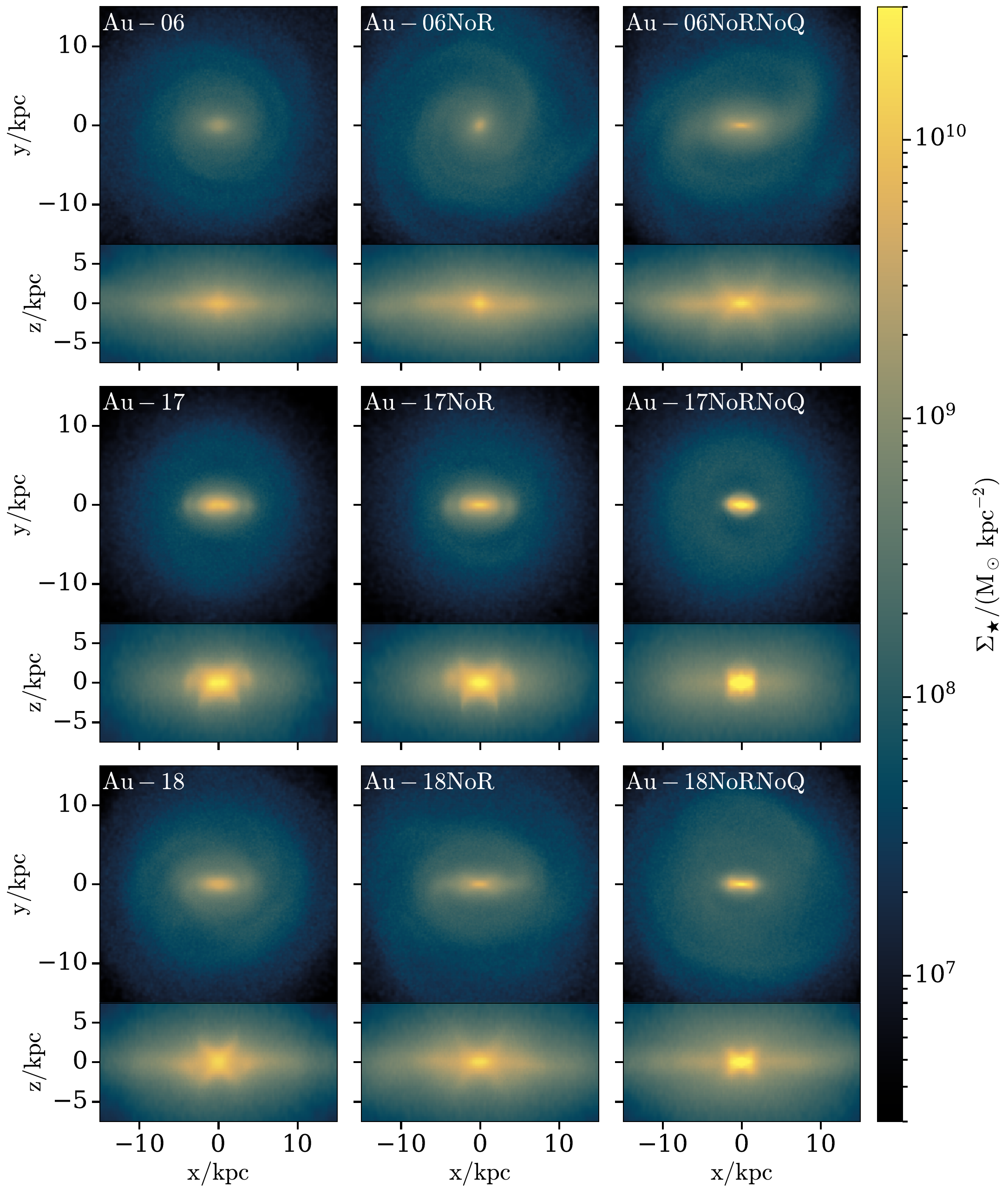}
\caption{Stellar surface density projection at $z$ = 0. The top, middle, and bottom rows contain results for Au-06, Au-17, and Au-18, respectively. The left-hand, middle, and right-hand columns contain results for the fiducial halo, the NoR, and the NoRNoQ variant, respectively. Each row contains a face-on (top panel) and an edge-on (bottom panel) projection. Prominent boxy/peanut bulges appear in most edge-on projections.}
\label{fig:stellar_density_combination}
\end{figure*}

\Fig{stellar_density_combination} shows the stellar surface density projection at $z$ = 0. The top, middle, and bottom rows contain results for Au-06, Au-17, and Au-18, respectively. The left-hand, middle, and right-hand columns contain results for the fiducial halo, the NoR, and the NoRNoQ variant, respectively. Each row contains a face-on (top panel) and an edge-on (bottom panel) projection.

A notable difference between each fiducial halo and its two variants is the distribution of stellar particles in the central regions which reflects dynamically distinct components. For example, we can see that removing AGN feedback from a weakly barred galaxy (Au-06) gives rise to the formation of a bar, which when viewed edge-on reveals a prominent peanut bulge \citep[see][for a review]{LS16}. This characteristic X-shaped feature has been studied in numerous observed \citep{WB88,MWH95,BAA06,LS17} and simulated galaxies \citep{CS81,CD90,PF91,MSH06} and its origin has been connected with either vertical resonance orbits with 2:1\footnote{Two vertical oscillations for every revolution in the bar frame.} frequency ratio \citep{Q02,QMS14,PSS20} or buckling instabilities \citep{RSJ91,MS94}. 

\begin{figure*}
\centering \includegraphics[width=\textwidth]{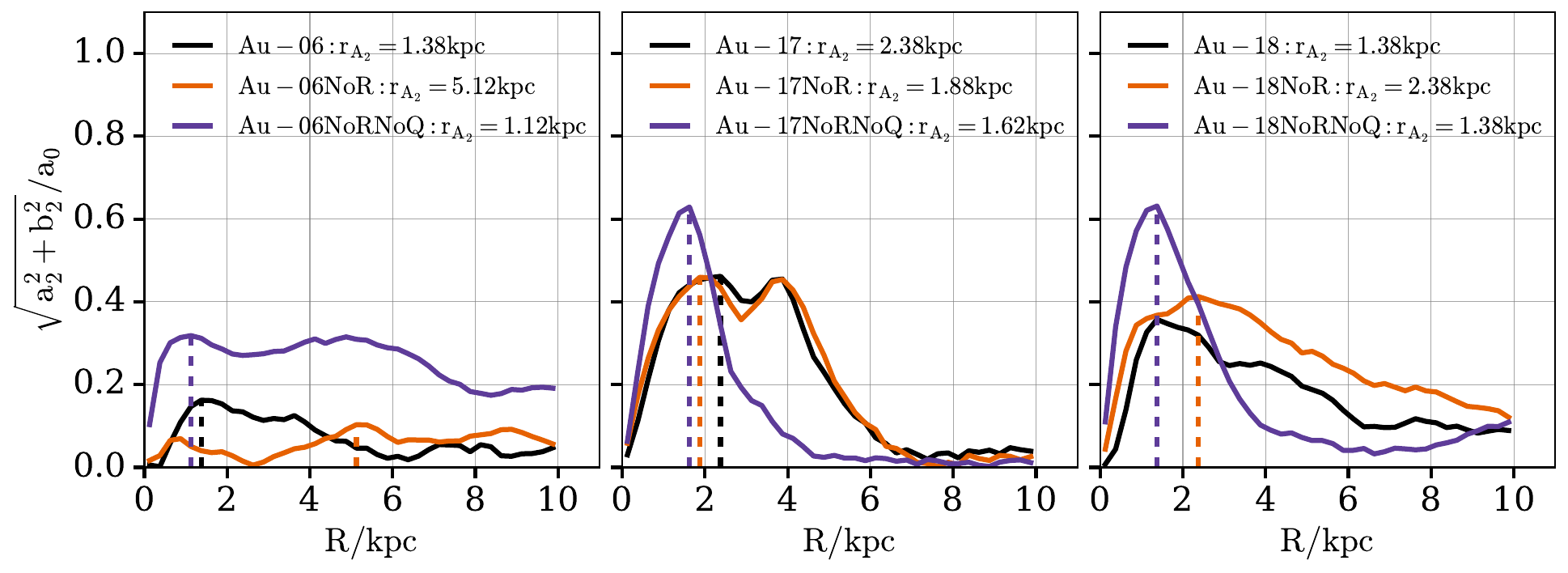}
\centering \includegraphics[width=\textwidth]{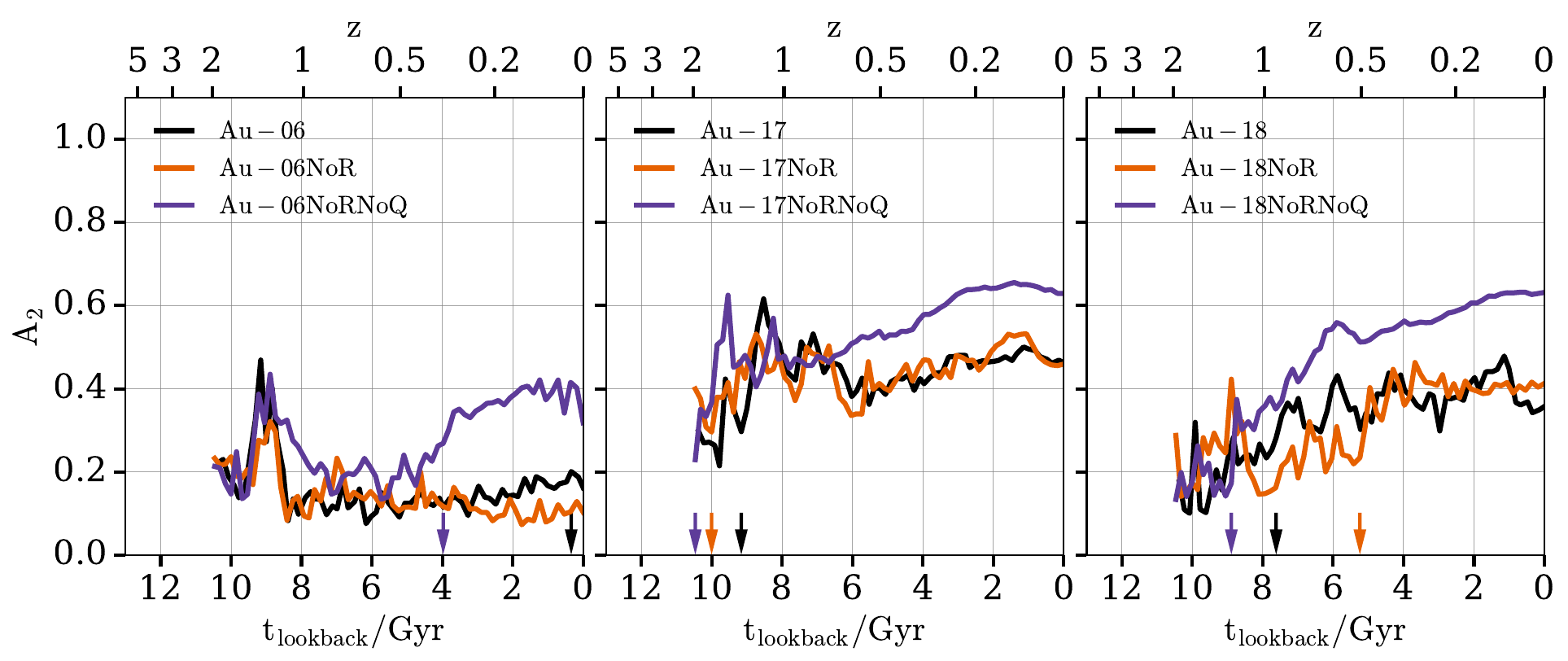}
\caption{Radial profile of the relative amplitude of the $m$ = 2 Fourier component at $z$ = 0 (top row) along with the evolution of the bar strength (bottom row). In each row the left-hand, middle, and right-hand panels contain results for Au-06, Au-17, and Au-18, respectively. The black, orange, and purple curves show the fiducial halo, the NoR, and the NoRNoQ variant, respectively. In the top row, the vertical dashed black, orange, and purple lines show the radius of the maximum of the $A_2$ profile for the fiducial halo, the NoR, and the NoRNoQ variant, respectively. In the bottom row, the black, orange, and purple arrows point at the bar formation time (see the text for more information) of the fiducial halo, the NoR, and the NoRNoQ variant, respectively. All NoRNoQ variants have stronger (i.e. higher $A_2$ values) and shorter bars than the fiducial and NoR variants and they develop them earlier.}
\label{fig:bar_strength_profile_combination}
\end{figure*}

To be able to quantitatively compare the bar strength between the different setups, we follow \cite{AMR13} and calculate the Fourier components of the face-on stellar density as
\begin{flalign} \label{eq:}
a_m(R) = \sum_{i=0}^N m_i\ \mathrm{cos}(m\theta_i)\; , \; \; \; \; m=0, 1,2,..., && \\
b_m(R) = \sum_{i=0}^N m_i\ \mathrm{sin}(m\theta_i)\; , \; \; \; \; m=0, 1,2,..., &&
\end{flalign}
where $R$ is the cylindrical radius, $m_i$ is the mass, $\theta_i$ is the azimuthal angle of stellar particle $i$, and the sum goes over all $N$ stellar particles. Hence, the bar strength is obtained from the maximum value of the $m$ = 2 relative Fourier component as
\begin{flalign} \label{eq:}
A_2 = \mathrm{max} \left( \frac{\sqrt{a_2^2(R) + b_2^2(R)}}{a_0(R)} \right) , &&
\end{flalign}
and in this work we define a strongly-barred galaxy as one having $A_2 \geq$ 0.3 \citep{FGP20} and a weakly-barred as one having 0.2 $\leq A_2$ < 0.3. We term Au-06 a very weakly barred galaxy since, as we show below, its bar strength reaches values just above 0.2. In addition, we visually check the morphology of each galaxy in order to ensure that a bar structure exists and that the $A_2$ values are not artificially high due to mergers/interactions.

\Fig{bar_strength_profile_combination} shows the radial profile of the relative amplitude of the $m$ = 2 Fourier component at $z$ = 0 (top row) along with the evolution of the bar strength (bottom row). In each row the left-hand, middle, and right-hand panels contain results for Au-06, Au-17, and Au-18, respectively. The black, orange, and purple curves show the fiducial halo, the NoR, and the NoRNoQ variant, respectively. In the top row, the vertical dashed black, orange, and purple lines show the radius of the maximum of the $A_2$ profile for the fiducial halo, the NoR, and the NoRNoQ variant, respectively. In the bottom row, the black, orange, and purple arrows point at the bar formation time (which we define as the time at which $A_2$ crosses the weak or strong bar threshold and never drops below it for more than two snapshots) of the fiducial halo, the NoR, and the NoRNoQ variant, respectively.

Au-17 and Au-18 -- which are both strongly barred galaxies -- show a significant increase on their bar strength when the quasar mode feedback is turned off (i.e. in their NoRNoQ variants), whilst the NoR variants show small deviations from the fiducial haloes. In addition, we see that all NoRNoQ variants form persistently strong bars before the corresponding fiducial halo and the NoR variant, hence the NoRNoQ bars are both stronger and form earlier, as indicated by the arrows in the bottom row of \Fig{bar_strength_profile_combination}. Therefore, \Fig{bar_strength_profile_combination} suggests that the lack of AGN quasar mode not only can enhance the strength of a bar (as is the case for Au-17 and Au-18) but it can also turn a very weakly-barred galaxy (Au-06) into a strongly-barred one (Au-06NoRNoQ).

While the bar strength is increased in the NoRNoQ variants, we find that the length of the bar in these runs decreases. There are several methods one can use to estimate the length of the bar \cite[e.g.][]{AM02}; here we use a proxy for the bar length the radius of the maximum relative $m$ = 2 Fourier component ($A_2$). While this method underpredicts the true length of bars \citep[e.g.][]{HMB20} in this work we are only interested in comparing the bar length between different variants of the same halo, and for such purposes the maximum of $A_2$ is sufficient. As can been seen by this method, all NoRNoQ runs produce shorter bars than their fiducial and NoR counterparts. However, as discussed by \cite{E05} and \cite{G11} larger galaxies contain larger bars. Hence, in order to check the robustness of this conclusion and any potential dependence of bar length on disc scale length or morphology \citep{AM80,EE85,M85}, in \Sec{Present day galactic properties:Stellar mass distribution} we provide an additional estimate of the bar length (based on photometric decomposition) where we normalise it by the galaxy size.

\begin{figure*}
\centering 
\includegraphics[width=\textwidth]{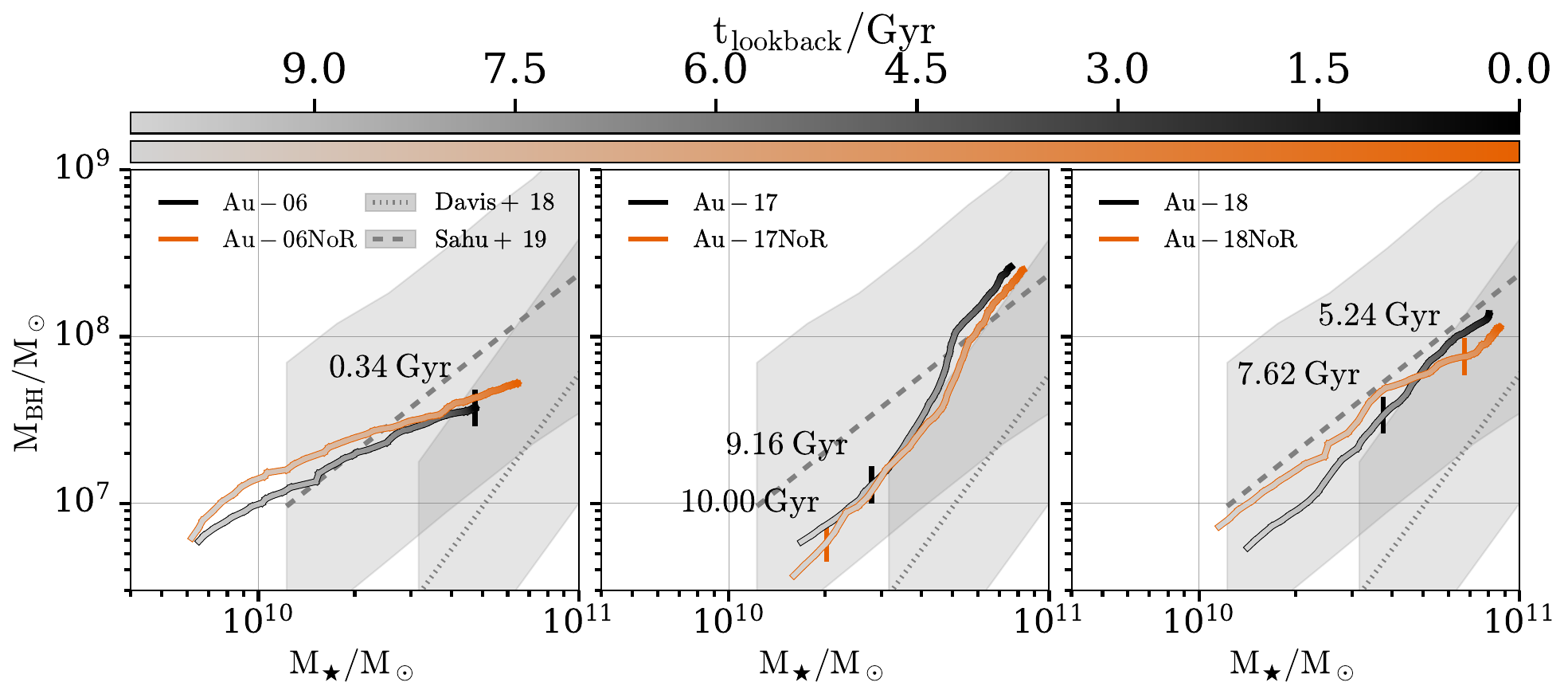}
\caption{Evolution of the black hole mass - stellar mass relation. The left-hand, middle, and right-hand panels contain results for Au-06, Au-17, and Au-18, respectively. The black and orange curves show results for the fiducial halo and the NoR variant, respectively. The black and orange vertical bar symbols indicate the bar formation time for the fiducial halo and the NoR variant, respectively. The grey dashed and dotted lines show the \protect\cite{SGD19} sample of local, early type barred galaxies and the \protect\cite{DGC18} sample of local spiral galaxies, respectively. The shaded regions show the 1$\sigma$ rms scatter in the data of the aforementioned observational studies. Galaxies with and without radio mode feedback follow similar paths on the black hole mass - stellar mass plane.}
\label{fig:stellar_blackhole_masses_combination}
\end{figure*}

In order to further investigate the co-evolution of bars and black holes, we explore the relation between bar formation and black hole mass evolution (see also \Sec{Quasar mode effects} and \Sec{Radio mode effects} for the evolution of the AGN feedback energy).

\Fig{stellar_blackhole_masses_combination} shows the evolution of the black hole mass - stellar mass relation. The left-hand, middle, and right-hand panels contain results for Au-06, Au-17, and Au-18, respectively. The black and orange curves show results for the fiducial halo and the NoR variant, respectively.\footnote{We remind the reader that the NoRNoQ variants are not included in this plot since their black hole particles have been completely removed.} The black and orange vertical bar symbols indicate the bar formation time for the fiducial halo and the NoR variant, respectively. The grey dashed and dotted lines show the \cite{SGD19} sample of local, early type barred galaxies and the \cite{DGC18} sample of local spiral galaxies, respectively. The shaded regions show the 1$\sigma$ rms scatter in the data of the aforementioned observational studies.

We see that both galaxies with AGN feedback and their NoR variants end up at $z$ = 0 in good agreement with observations. In addition, we find no obvious relation between black hole mass and bar formation time, in agreement with past theoretical and observational studies \citep[e.g.][]{GMG17,RBD20}. However, as we further discuss in \Sec{Discussion}, we intend to explore in depth how bars affect the growth of black holes in \auriga\ galaxies, in a future study.

\subsection{Stellar-halo mass relation} \label{sec:Present day galactic properties:Stellar-halo mass relation}

\begin{figure}
\centering \includegraphics[width=0.49\textwidth]{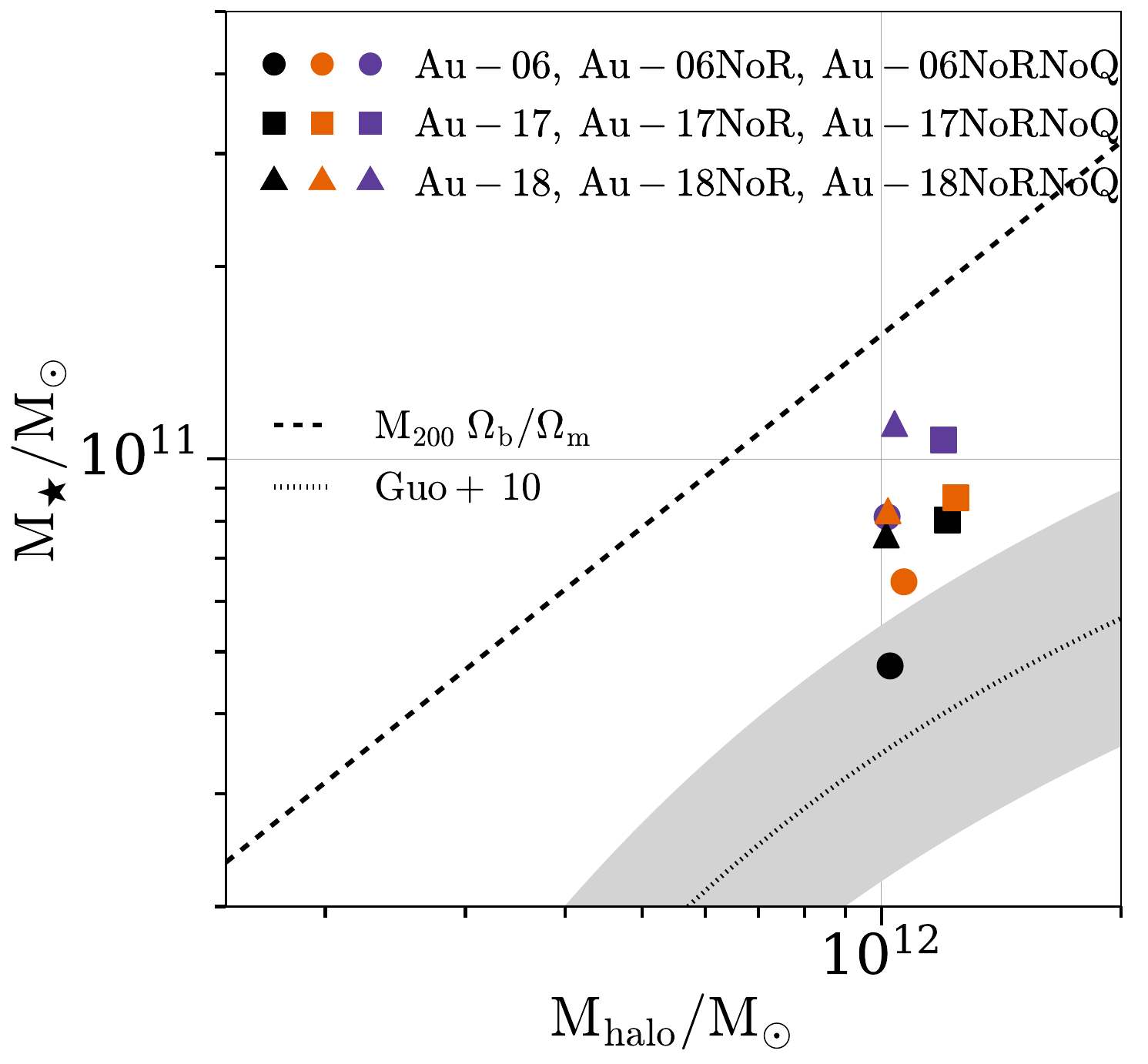}
\caption{Stellar mass as a function of the halo mass at $z$ = 0. The former is defined as the sum over all stellar particles within 0.1$R_{200}$ and the latter as the sum over all mass elements within $R_{200}$. The circles, triangles, and squares represent Au-06, Au-17, and Au-18, respectively. The black, orange, and purple colours represent the fiducial halo, the NoR, and the NoRNoQ variant, respectively. The dashed diagonal line shows the baryon conversion efficiency (see the text for more information) and the dotted curve the stellar-halo mass relation from \protect\cite{GWL10}. For all galaxies the NoRNoQ variant has a more massive stellar component than the NoR, which in turn is more massive than the fiducial galaxy.}
\label{fig:stellar_vs_halo_mass_combination}
\end{figure}

\Fig{stellar_vs_halo_mass_combination} shows the stellar mass as a function of the halo mass at $z$ = 0. The former is defined as the sum over all stellar particles within 0.1$R_{200}$ and the latter as the sum over all mass elements within $R_{200}$. The circles, triangles, and squares represent Au-06, Au-17, and Au-18, respectively. The black, orange, and purple colours represent the fiducial halo, the NoR, and the NoRNoQ variant, respectively. The dashed diagonal line shows the baryon conversion efficiency \citep{GWL10,MNW13} with \Omegam\ = 0.307 and \Omegab\ = 0.048 \citep{PC14,GGM17}, and the dotted curve the stellar-halo mass relation from \cite{GWL10}. 

As expected, there is a clear trend which shows that all fiducial haloes have lower stellar masses than their NoR variants which in turn have lower stellar masses than their NoRNoQ variants. Arguably, the NoR and NoRNoQ variants move (almost) vertically in the stellar-halo mass plane hence these galaxies end up further away from the abundance matching predictions \citep{BCW10,MSM10}. It is worth noting that, as has been already reported by \cite{GGM17} (see their Section 5.3), even the fiducial haloes lie above the abundance matching relation, since the AGN feedback can be insufficient in suppressing early star formation. However, having galaxies more massive than what the abundance matching relation dictates can in fact lead to bars which rotate fast (i.e. have a corotation/bar length ratio less than 1.4) which is more consistent with what is found in observations \citep{FGP21}.

Investigating how this additional stellar mass in the NoR and NoRNoQ variants (compared to the fiducial haloes) is distributed in the galaxy is essential in order to understand its dynamical and structural properties. It is well known that massive discs are prone to disc instabilities and bar formation \citep{ELN82,MMW98,YS15,ITH19}. On the other hand, a substantial central component tends to delay or even suppress bar formation \citep{A04,A13,KD19}. In order to identify which of the above two mechanisms is the dominant one in our galaxies, we analyse below the distribution of stellar mass and calculate the relative contribution of the bar/bulge/disc to the total stellar mass by performing a 2D three component decomposition.

\subsection{Stellar mass distribution} \label{sec:Present day galactic properties:Stellar mass distribution}

\begin{figure*}
\centering \includegraphics[width=\textwidth]{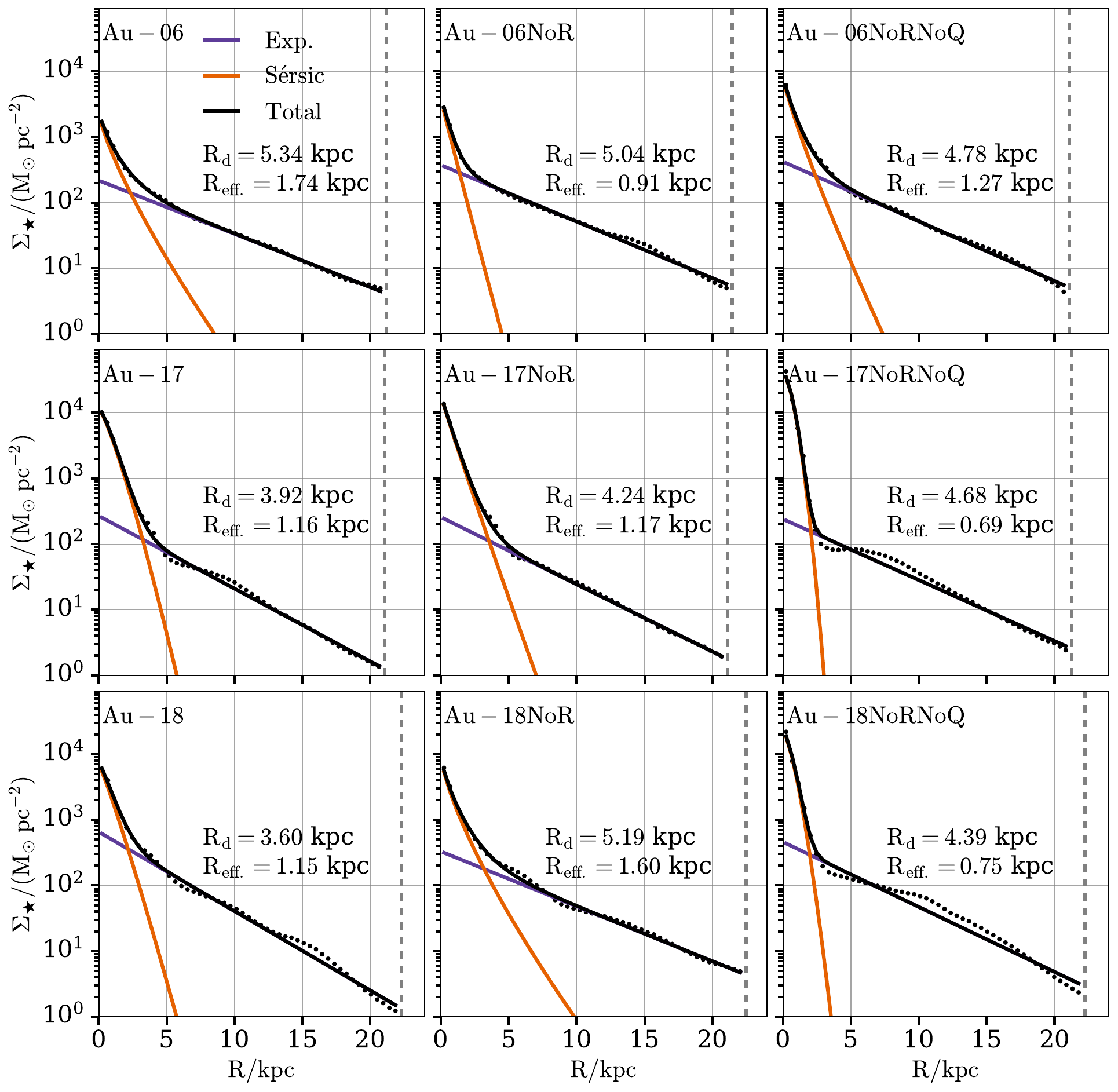}
\caption{Radial stellar surface density profiles at $z$ = 0. The top, middle, and bottom rows contain results for Au-06, Au-17, and Au-18, respectively. The left-hand, middle, and right-hand columns contain results for the fiducial halo, the NoR, and the NoRNoQ variant, respectively. We only consider stellar particles within 0.1$R_{200}$ along the disc plane (vertical dashed grey line) and $\pm$ 5 kpc in the vertical direction. The total fitted profile (black curve) is a combination of a S\'ersic (orange curve) and exponential (purple curve) profile and it was carried out using a non-linear least squares method \protect\citep{MPS14,GGM17}. The NoR and NoRNoQ variants have higher concentration of stellar mass in the central regions compared to their fiducial halo.}
\label{fig:stellar_surface_density_profiles_combination}
\end{figure*}

\begin{table*}
	\centering
	\caption{The fitting parameters from \Fig{stellar_surface_density_profiles_combination}. The rows represent 1) model names (three fiducial runs, namely Au-06, Au-17, and Au-18, plus a no-radio and a no-radio-no-quasar variant for each one); 2) total stellar mass as defined in \Fig{stellar_surface_density_profiles_combination}; 4) S\'ersic index; 5) effective radius; 6) inferred S\'ersic mass; 8) scale length; 			9) inferred disc mass; 10) disc-to-total stellar mass ratio.}
	\label{tab:stellar_surface_density_profiles_combination_table}
	\begin{tabular}{lccccccccc} 
		\hline
		  & Au-06 & Au-06NoR & Au-06NoRNoQ & Au-17 & Au-17NoR & Au-17NoRNoQ & Au-18 & Au-18NoR & Au-18NoRNoQ \\
		\hline
		\Mstar/(10$^{10}$ \Msun) & 4.57 & 6.24 & 7.94 & 7.38 & 8.12 & 11.14 & 7.77 & 8.35 & 10.35 \\
		\textbf{S\'ersic} & & & & & & & & & \\
        $n$ & 1.28 & 1.02 & 1.17 & 0.81 & 1.05 & 0.59 & 0.90 & 1.35 & 0.70 \\
        \Reff/kpc & 1.74 & 0.91 & 1.27 & 1.16 & 1.17 & 0.69 & 1.15 & 1.60 & 0.75 \\
        \MSersic/(10$^{10}$ \Msun) & 1.00 & 0.67 & 2.24 & 4.68 & 5.22 & 7.53 & 2.40 & 3.00 & 4.36 \\
		\textbf{Exponential} & & & & & & & & & \\
	    $h$/kpc& 5.34 & 5.04 & 4.78 & 3.92 & 4.24 & 4.68 & 3.60 & 5.19 & 4.39 \\
        \Md/(10$^{10}$ \Msun) & 3.89 & 5.95 & 6.01 & 2.58 & 2.89 & 3.29 & 5.31 & 5.58 & 5.55 \\
        $D/T$ & 0.80 & 0.90 & 0.73 & 0.36 & 0.36 & 0.30 & 0.70 & 0.65 & 0.56 \\
		\hline
	\end{tabular}
\end{table*}

\Fig{stellar_surface_density_profiles_combination} shows the radial stellar surface density profiles at $z$ = 0. The top, middle, and bottom rows contain results for Au-06, Au-17, and Au-18, respectively. The left-hand, middle, and right-hand columns contain results for the fiducial halo, the NoR, and the NoRNoQ variant, respectively. We only consider stellar particles within 0.1$R_{200}$ along the disc plane (vertical dashed grey line) and $\pm$ 5 kpc in the vertical direction. The total fitted profile (black curve) is a combination of a S\'ersic (orange curve) and exponential (purple curve) profile and it was carried out using a non-linear least squares method \protect\citep{MPS14,GGM17}.

The steepening of the stellar surface density profiles in the very centre as we move from the fiducial to the NoR and NoRNoQ variants indicates the higher concentration of mass in the centre. This behaviour reveals that there is an excess of central mass in these variants which is not present in the fiducial haloes. The fitting data appearing in \Tab{stellar_surface_density_profiles_combination_table} further illustrates this point. As shown in \Sec{Present day galactic properties:Stellar-halo mass relation}, when either the radio mode or both AGN modes are turned off, the resulting galaxy becomes more massive, as expected. This is reflected in the increase of the stellar mass inferred both from the S\'ersic (sixth row) and the exponential (ninth row) profiles, respectively. However, the relative contribution of the disc to the total stellar mass ($D/T$) presented in the last row shows that all NoRNoQ variants' $D/T$ ratios decrease. In other words, their central components (i.e. the ones fitted with a S\'ersic profile) not only become more massive but also have higher relative contributions to the total stellar mass (see also \App{Stellar surface density profiles}). At the same time these components also decrease their effective radii. Hence, the lack of AGN feedback results in both more massive and more concentrated central distributions of mass. A similar conclusion is hard to draw for the NoR runs since they do not seem to have a consistent behaviour (Au-06NoR, Au-17NoR, and Au-18NoR have, respectively, higher, the same, and lower $D/T$ with respect to their fiducial haloes).

Regarding the disc components (i.e. the ones fitted with an exponential profile), we find that the scale lengths of Au-17 and Au-18 -- which are both strongly barred galaxies -- increase when the radio mode feedback is turned off since their circumgalactic medium (CGM) gas can cool more efficiently and settle in a more extended disc. A similar trend was found by \cite{GGM17} (see their Section 4.3) who re-simulated Au-22 with its AGN feedback turned off after $z$ = 1, and showed that AGN feedback can suppress the formation of a large stellar disc. However, we see that Au-06 -- our very weakly barred galaxy -- does not follow the same trend as its NoR variant has a smaller disc (see also discussion below).

In order to understand the effect of AGN feedback on the structure of MW-mass galaxies, it is important to identify in which components these additional stellar particles belong to. Do they form a classical or a pseudo-bulge, or are they part of a bar? To address this question, we use \imfit\footnote{\href{https://www.mpe.mpg.de/~erwin/code/imfit/}{https://www.mpe.mpg.de/$\sim$erwin/code/imfit/}} \citep{E15} to perform a 2D bar/bulge/disc decomposition and quantify the change in the relative contribution of each component between the different variants. Since our intention is to compare the fiducial \auriga\ haloes with their NoR and NoRNoQ variants, we use images that accurately represent the mass distribution, not necessarily ones that mimic observations. Therefore, we fit each galaxy with a combination of the following profiles (see \Tab{Imfit}):
\begin{enumerate}[wide=0pt, label=(\roman*)]
\item An elliptical 2D exponential function for the stellar disc component, with the major-axis intensity profile given by
\begin{flalign} \label{eq:Exponential}
I(\alpha) = I_0\ \exp\,( -\alpha/h) \;, &&
\end{flalign}
where $I_0$ is the central surface brightness, $\alpha$ is the distance from the image centre, and $h$ is the scale length.

\item A 2D analogue of the Ferrers ellipsoid for the bar component where the intensity profile is given by
\begin{flalign} \label{eq:Ferrers}
I(m) = I_0(1-m^2)^{n} \;, &&
\end{flalign}
where $n$ controls the sharpness of the truncation at the end of bar and $m^{2}$ is defined as
\begin{flalign} \label{eq:m_squares}
m^2 = \left( \frac{|x|}{\alpha} \right)^{c_0 +2} \left( \frac{|y|}{b} \right)^{c_0 +2} \;, &&
\end{flalign}
where $x$ and $y$ describe the position on the image, $\alpha$ and $b$ are the semi-major and semi-minor axes, respectively, and $c_0$ defines the shape of the isophotes \citep[zero, negative, and positive values result in pure ellipses, disky, and boxy ellipses, respectively; see][]{AMW90}. We note that the intensity is constant on the ellipses and goes to zero outside a specific semi-major axis (sma) value (\abar).

\item An elliptical 2D S\'ersic function for the bulge component with the major-axis intensity profile given by
\begin{flalign} \label{eq:Sersic}
I(\alpha) = \Ie\ \exp\left\lbrace - \bn \left[ \left( \frac{\alpha}{\re} \right)^{1/n} -1 \right] \right\rbrace \;, &&
\end{flalign}
where \Ie\ is the surface brightness at the half-light radius (\re), $n$ is the S\'ersic index, and \bn\ is calculated for $n >$ 0.36 via the polynomial approximation of \cite{CB99} and the approximation of  \cite{MCH03} for $n \leq$ 0.36.
\end{enumerate}

\begin{table*}
	\centering
	\caption{The fitting parameters from \imfit. The rows represent 1) model names (three fiducial runs, namely Au-06, Au-17, and Au-18, plus a no-radio and a no-radio-no-quasar variant for each one); 3) scale length; 4) disc/total fraction; 6) S\'ersic index; 7) effective radius; 8) bulge/total fraction; 10) ellipticity; 11) bar length; 12) bar/total fraction. Horizontal bar symbols indicate that the corresponding profile was not used in the corresponding galaxy.}
	\label{tab:Imfit}
	\begin{tabular}{lccccccccc} 
		\hline
		  & Au-06 & Au-06NoR & Au-06NoRNoQ & Au-17 & Au-17NoR & Au-17NoRNoQ & Au-18 & Au-18NoR & Au-18NoRNoQ \\
		\hline
		\textbf{Exponential} & & & & & & & & & \\
	    h/kpc & 8.20 & 9.61 & 7.99 & 5.11 & 4.69 & 5.04 & 6.85 & 8.50 & 7.88 \\
		Fraction & 0.96 & 0.98 & 0.95 & 0.89 & 0.87 & 0.97 & 0.93 & 0.97 & 0.98 \\
		\textbf{S\'ersic} & & & & & & & & & \\
        $n$ & - & 0.79 & - & - & - & - & - & - & - \\
	    \re /kpc & - & 2.58 & - & - & - & - & - & - & - \\
		Fraction & - & 0.02 & - & - & - & - & - & - & -\\
		\textbf{FerrersBar2D} & & & & & & & & & \\
		Ellipticity & 0.60 & - & 0.59 & 0.55 & 0.60 & 0.60 & 0.60 & 0.69 & 0.66 \\
	    \abar/kpc & 4.69 & - & 6.56 & 6.09 & 6.09 & 3.63 & 6.10 & 7.03 & 3.98 \\
  	    \abar/h & 0.57 & - & 0.82 & 1.19 & 1.30 & 0.72 & 0.89 & 0.82 & 0.51 \\
		Fraction & 0.04 & - & 0.05 & 0.10 & 0.13 & 0.03 & 0.07 & 0.03 & 0.02 \\
		\hline
	\end{tabular}
\end{table*}

The decomposition parameters are presented in \Tab{Imfit} and the corresponding plots in \App{2D decompositions}. The only galaxy with a prominent round central component (low ellipticity) is Au-06NoR (see \Fig{Au-06}). Hence, we choose an exponential plus a S\'ersic profile to accurately model this galaxy; for the remaining galaxies we use an exponential plus a Ferrers profile since they do not appear to have a central spheroidal component. In general, our results are in agreement with the study of \cite{BFP20}. However, we find systematically larger disc scale lengths (by a factor of 1.2-1.5) that may be due to the slightly different methodologies employed in producing the images to be fitted and in the fitting itself. For example, \cite{BFP20} fit all galaxies in their sample with three components (exponential+S\'ersic+Ferrers).

As already mentioned in \Sec{Present day galactic properties:Morphological properties}, Au-06 is a weakly barred galaxy which becomes a strongly barred one (i.e. $A_2$ > 0.3) when its AGN feedback is absent. This finding is further supported by its 2D decomposition, since we see that Au-06NoRNoQ's bar is longer and has higher contribution to the total galaxy mass compared to Au-06. Furthermore, even though Au-06NoRNoQ is more massive than Au-06 (see \Fig{stellar_vs_halo_mass_combination}) we find that the former has smaller disc scale length than the latter. Hence, we report that when the AGN feedback of our weakly barred galaxy is turned off it results in a galaxy with longer and stronger bar (see \Fig{bar_strength_profile_combination}) while its stellar disc shrinks.\footnote{From \Tab{stellar_surface_density_profiles_combination_table} we drew the same conclusion for Au-06NoR, however \Tab{Imfit} suggests the opposite -- the stellar disc of Au-06NoR is larger than that of Au-06. This disagreement may be due to the different profiles used for Au-06 (S\'ersic+exponential in \Tab{stellar_surface_density_profiles_combination_table} and Ferrers+exponential in \Tab{Imfit}) since both decomposition methods are sensitive to parameters such as how well the inner components are fit and/or how far out the bulge and the disc extend.}

On the contrary, for Au-17 and Au-18 (which are both strongly barred galaxies) we find that both the absolute (\abar) and the normalised (\abar/$h$) bar lengths of the NoRNoQ variants are shorter than those of the fiducial haloes. Moreover, these variants also have lower $Bar/T$ ratios compared to their fiducial runs. In addition, we see that the absence of AGN feedback results in more extended discs which are also more massive, as reported in \Sec{Present day galactic properties:Stellar-halo mass relation}.

In summary, we find that Au-06NoRNoQ develops a bar which is longer than Au-06's, in contrast to Au-17NoRNoQ and Au-18NoRNoQ which form shorter bars than their fiducial haloes. Furthermore, we report that by completely removing the AGN feedback our galaxies become more prone to forming strong bars (i.e. with $A_2$ > 0.3), thus the increased concentration of stellar mass in the centres of the NoRNoQ variants in \Fig{stellar_surface_density_profiles_combination} does not represents a classical bulge-like population (see also \Sec{Discussion}). These results reveal a significant effect of ejective AGN feedback on the structure of MW-mass galaxies which has been previously thought to be negligible \citep{CAR16,RVC16,ZL16}. This effect is particularly important for large-scale simulations, since different AGN implementations can change the morphology and dynamics of galaxies and consequently the corresponding statistical predictions \citep[e.g. the fraction of barred galaxies in a particular simulated volume][]{ANA17,PL19,RBD20}.

\section{Quasar mode effects} \label{sec:Quasar mode effects}

As we saw in the previous section, the different AGN feedback variants give rise to different structural and dynamical properties. In this section, we explore how the quasar mode feedback affects our galaxies' star formation rate histories (\Sec{Quasar mode effects:Star formation rate histories}) and gas properties (\Sec{Quasar mode effects:Gas properties}).

\subsection{Star formation rate histories} \label{sec:Quasar mode effects:Star formation rate histories}

\begin{figure*}
\centering \includegraphics[width=\textwidth]{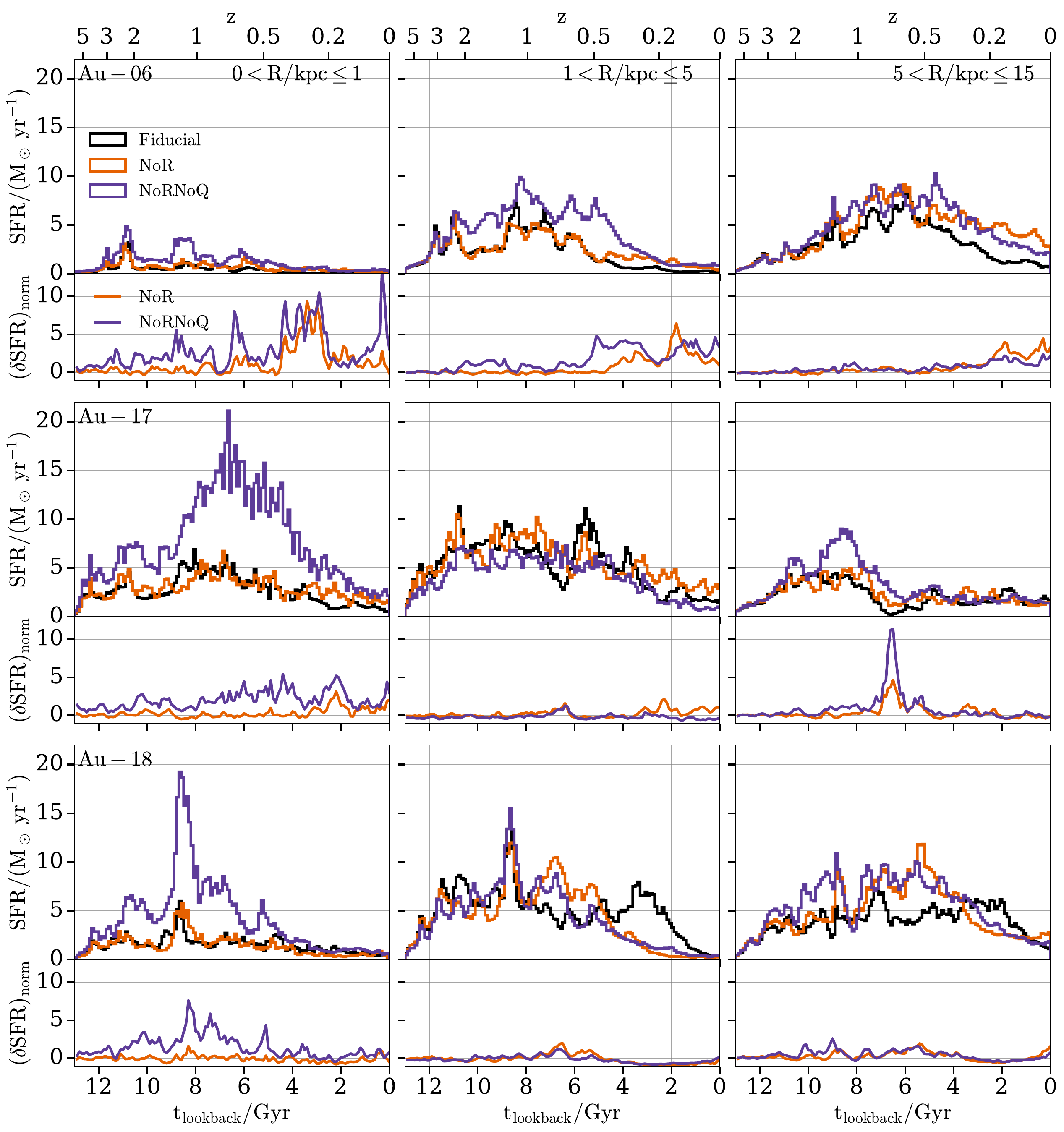}
\caption{Star formation rate histories along with the normalised SFR difference between a variant and the fiducial halo (see the text for more information). The top, middle, and bottom rows contain results for Au-06, Au-17, and Au-18, respectively. The left-hand, middle, and right-hand columns contain results inside spherical apertures with radii 0 < $R$/kpc $\leq$ 1, 1 < $R$/kpc $\leq$ 5, and 5 < $R$/kpc $\leq$ 15, respectively. Each row contains the star formation rate histories (top panel) and the normalised SFR differences (bottom panel). The black, orange, and purple lines represent the fiducial, the NoR, and the NoRNoQ variant, respectively. The AGN feedback suppresses star formation more in the centre than in the disc.}
\label{fig:delta_sfr_regimes_combination}
\end{figure*}

\Fig{delta_sfr_regimes_combination} shows the star formation rate histories along with the normalised SFR difference between a variant and the fiducial halo defined as
\begin{flalign} \label{eq:dSFRnorm}
\dSFRnorm \equiv \frac{\SFRvariant - \SFRfiducial}{\SFRfiducial} \;. &&
\end{flalign}
The top, middle, and bottom rows contain results for Au-06, Au-17, and Au-18, respectively. The left-hand, middle, and right-hand columns contain results inside spherical apertures with radii 0 < $R$/kpc $\leq$ 1, 1 < $R$/kpc $\leq$ 5, and 5 < $R$/kpc $\leq$ 15, respectively. Each row contains the star formation rate histories (top panel) and the normalised SFR differences (bottom panel). The black, orange, and purple lines represent the fiducial, the NoR, and the NoRNoQ variant, respectively.

As expected the NoRNoQ variants have on average the highest star-formation rates since the removed AGN feedback allows the gas to cool quicker and form stars more efficiently \citep[see also the third row of fig. 17 of][]{GGM17}. This effect is more prominent near the centre (0 < $R$/kpc $\leq$ 1) for all haloes, which shows that the quasar mode has a significant effect on star formation in the inner regions. In the outer disc ($R$/kpc > 1), the NoR and NoRNoQ variants follow in most cases similar trends, suggesting that the quasar AGN feedback mode has lower impact on the star formation rates away from the centre compared to the radio mode. 

However, AGN feedback is not the only mechanism that can affect star formation (see also \Sec{Discussion}). As seen in \Sec{Present day galactic properties:Morphological properties}, all NoRNoQ variants have the strongest bars throughout their lifetime and multiple studies that analysed the effects of bars on star formation concluded that (especially strong) bars can have an impact on central star formation \citep{GCD15,FAB16,SBD17,KHD18}. Bars drive gas towards the centre due to their non-axisymmetric potential which exerts strong torques on the gas \citep{SFB89,PM06,CDE14}. Consequently, as this gas accumulates at the centre it triggers the formation of stars, and the stronger the bar the more centrally concentrated the star formation \citep{A92,JSK05,WAY20}. 

\subsection{Gas properties} \label{sec:Quasar mode effects:Gas properties}

\begin{figure*}
\centering \includegraphics[width=\textwidth]{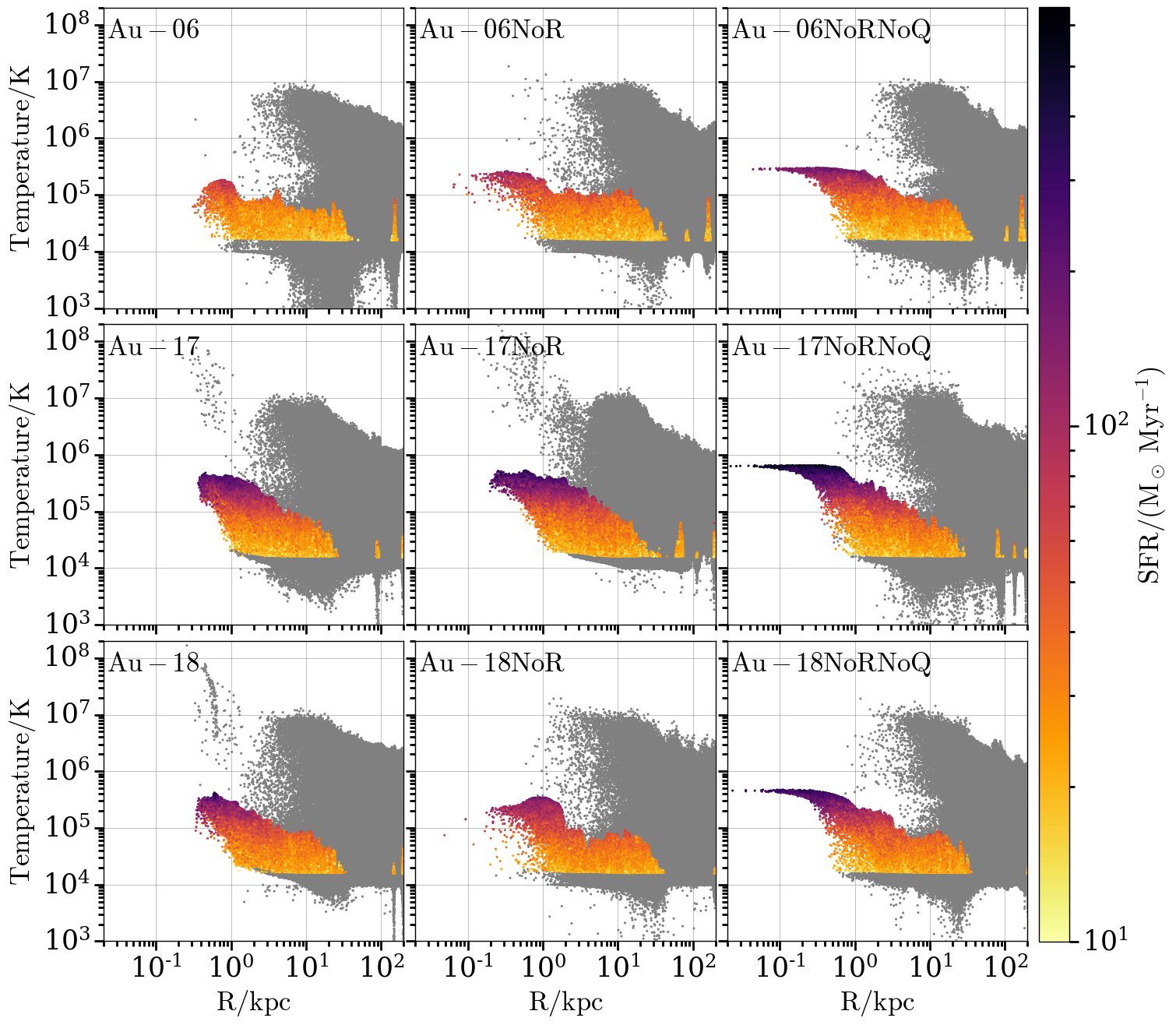}
\caption{Gas temperature as a function of galactocentric distance at $z$ = 0.05 (see also \App{Evolution of the gas temperature-distance relation}) colour-coded by the star formation rate of each gas cell (non-star-forming gas cells appear grey). The top, middle, and bottom rows contain results for Au-06, Au-17, and Au-18, respectively. The left-hand, middle, and right-hand columns contain results for the fiducial halo, the NoR, and the NoRNoQ variant, respectively. The quasar mode AGN feedback expels gas cells and reduces star formation rates in the central regions.}
\label{fig:gas_temperature_vs_distance_combination_005}
\end{figure*}

In the previous subsection, we saw that the quasar mode feedback reduces the SFR in the central kiloparsec. Here, we explore how the quasar mode affects star forming gas cells that surround the black hole.

\Fig{gas_temperature_vs_distance_combination_005} shows the gas temperature as a function of galactocentric distance at $z$ = 0.05 (see also \App{Evolution of the gas temperature-distance relation}) colour-coded by the star formation rate of each gas cell (non-star-forming gas cells appear grey). The top, middle, and bottom rows contain results for Au-06, Au-17, and Au-18, respectively. The left-hand, middle, and right-hand columns contain results for the fiducial halo, the NoR, and the NoRNoQ variant, respectively.

In all fiducial haloes and NoR variants there is an almost complete lack of gas cells at the very centre. On the other hand, all NoRNoQ variants not only have central gas cells but in all cases these cells also have the highest star formation rates, and form a characteristic horn-like feature on the gas temperature-distance plane. Furthermore, we see that the inner region ($R$ < 1 kpc) of all fiducial haloes and NoR variants is typically populated by very hot ($T$ > 10$^6$ K), non-star-forming gas cells which are not present in the NoRNoQ variants. These two findings show that the quasar mode AGN feedback heats up and expels the gas cells near the centre, thus it is responsible for suppressing the central star formation.

In order to better understand the interplay between the star forming gas cells and the quasar feedback mode, we analyse the evolution of the effective quasar mode energy (i.e. the feedback energy that is able to be absorbed by gas cells.\footnote{Star forming gas cells follow an effective equation of state (eEoS) which describes a cell's thermodynamical properties by connecting its pressure to its density \citep{SH03}. As described in \Sec{The Auriga simulations:Black hole and AGN feedback model} the AGN feedback in the \auriga\ simulations has two modes which both inject energy to gas cells. However, this energy will only be absorbed by cells that cool sufficiently i.e. that are not in the star-forming regime, while cells that are on the eEoS (i.e. have non-zero star forming rates), will instantaneously radiate away any thermal energy injection \citep{WSP18}.})

\begin{figure*}
\centering 
\includegraphics[width=\textwidth]{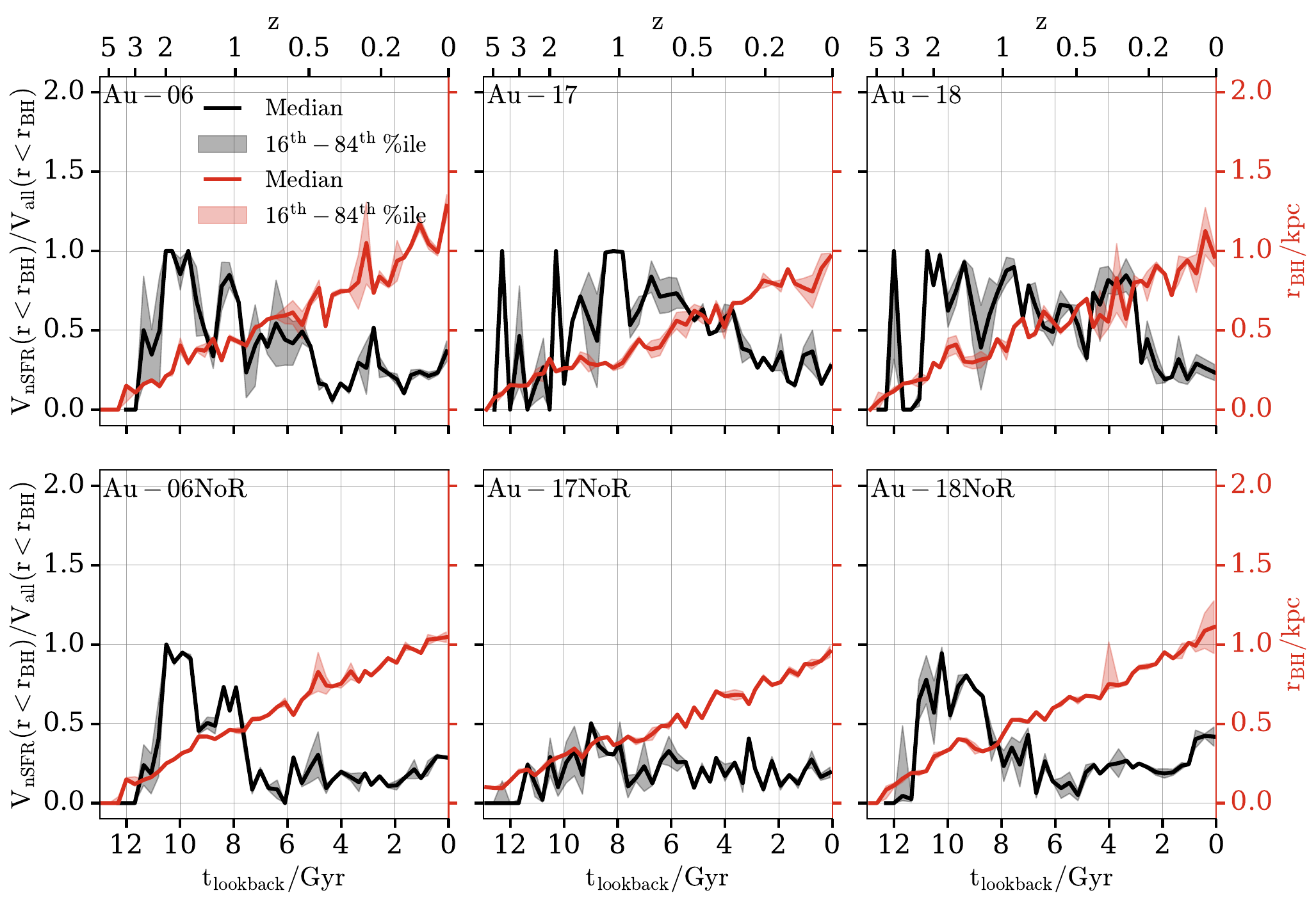}
\caption{Evolution of the ratio (black curve) between the volume occupied by all gas cells and that by the non-star-forming ones inside a sphere with radius \rBH\ (red curve). The left-hand, middle, and right-hand columns contain results for Au-06, Au-17, and Au-18, respectively. Each column contains on the top and bottom row the fiducial halo and the NoR variant, respectively. The black and red curves and shaded regions show the median and 16th–84th percentile range, respectively. The region around the black hole is populated by more non-star-forming cells when the AGN feedback is present.}
\label{fig:AGN_feedback_kernel_combination}
\end{figure*}

\Fig{AGN_feedback_kernel_combination} shows the evolution of the ratio (black curve) between the volume occupied by all gas cells and that by the non-star-forming ones inside a sphere with radius \rBH\footnote{\rBH\ is the physical radius enclosing the 384 $\pm$ 48 nearest gas cells around the black hole. We remind the reader that the quasar mode energy is injected in all cells which are within a sphere with radius \rBH.} (red curve). The left-hand, middle, and right-hand columns contain results for Au-06, Au-17, and Au-18, respectively. Each column contains on the top and bottom row the fiducial halo and the NoR variant, respectively. The black and red curves and shaded regions show the median and 16th–84th percentile range, respectively.

We can see that at various times there is a significant fraction of non-star-forming cells within the black hole's vicinity (i.e. within \rBH) which allow the quasar mode feedback to be effective. In the \auriga\ model, black holes eject energy before the cooling is treated and if cooling is inefficient to cool a gas cell down to the eEoS in one timestep, its star-formation rate is set to zero. Hence, the existence of non-star-forming cells indicates that during the next timestep the black hole feedback energy will be absorbed by those cells and not immediately get radiated away. For that reason, the ratio between the volume occupied by all gas cells and that by the non-star-forming ones can be used as a proxy to estimate the effectiveness of the quasar mode feedback in the \auriga\ simulations.

\begin{figure*}
\centering \includegraphics[width=\textwidth]{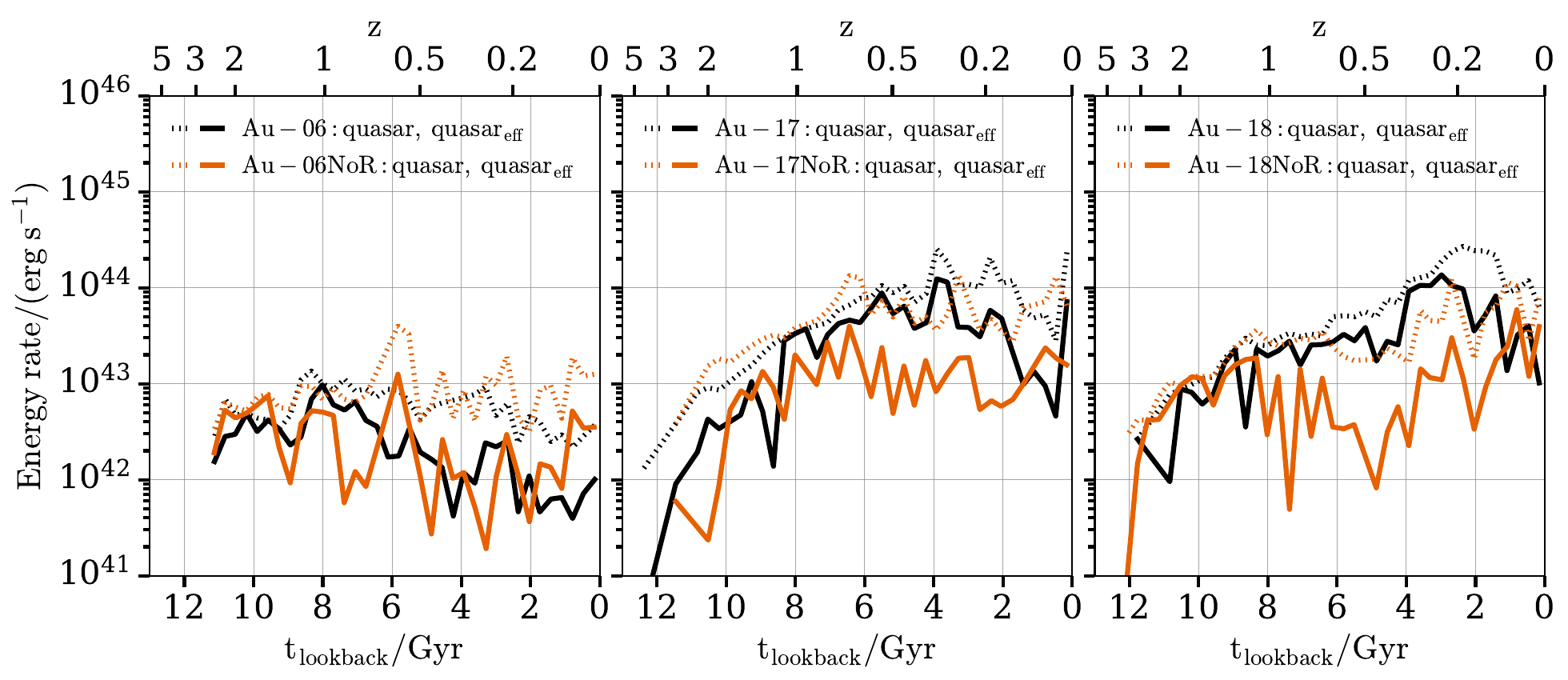}
\caption{Evolution of the quasar mode feedback energy rate. The left-hand, middle, and right-hand panels contain results for Au-06, Au-17, and Au-18, respectively. The black and orange curves show results for the fiducial halo and the NoR variant, respectively. The dashed curves show the total quasar mode energy as calculated by the model and the solid curves show the effective quasar mode energy (see the text for more information). For our strongly barred galaxies the quasar mode becomes on average less effective when the radio mode is absent.}
\label{fig:quasar_mode_distribution_combination}
\end{figure*}

\Fig{quasar_mode_distribution_combination} shows the evolution of the quasar mode feedback energy rate. The left-hand, middle, and right-hand panels contain results for Au-06, Au-17, and Au-18, respectively. The black and orange curves show results for the fiducial halo and the NoR variant, respectively. The dashed curves show the total quasar mode energy as calculated by the model (see \Sec{The Auriga simulations:Black hole and AGN feedback model} for more details) and the solid curves show the effective quasar mode energy. We define the effective quasar mode energy as the energy which is able to couple to the gas cells, i.e. which is not lost immediately due to radiative cooling by gas cells on the eEoS. We obtain this effective quasar mode energy by multiplying the total quasar mode energy with the volume ratios of non-star forming gas-to-total gas within a sphere with radius \rBH, presented in \Fig{AGN_feedback_kernel_combination}.

For Au-06 the quasar mode energy rate calculated by the model has an almost flat evolution with time while the effective one decreases for $z$ < 1 and becomes roughly an order of magnitude lower than the calculated. This trend holds both for the fiducial halo and the NoR variant. On the other hand, Au-17's and Au-18's quasar mode energy rates increase as we go to lower redshifts and are systematically higher than that of their NoR variants. Hence, we see that when the radio mode of our strongly barred galaxies is turned off, the quasar mode becomes on average less effective because the volume around the black hole is now occupied more by star forming gas cells (see \Fig{AGN_feedback_kernel_combination}). This indicates that the radio mode has an indirect impact on the gas in the inner disc via influencing the effectiveness of the quasar mode.

In conclusion, we report here that the quasar mode feedback has a significant effect on the gas in the inner regions of MW-mass galaxies. Even though it is not capable of fully quenching our galaxies \citep[as it does for massive ETGs,][]{MBT09,RTP19}, it has a significant impact on their morphology and dynamics.

\section{Radio mode effects} \label{sec:Radio mode effects}

Having presented the effects of the quasar mode feedback on galactic properties, in this section, we explore how the radio mode feedback affects galaxies in the \auriga\ simulations. We remind the reader that the radio mode feedback in \auriga\ is implemented in a way in which it affects the CGM of the galaxy by stochastically inflating bubbles at random locations to heat up the gas.

\begin{figure*}
\centering 
$\vcenter{\hbox{\includegraphics[width=0.49\textwidth]{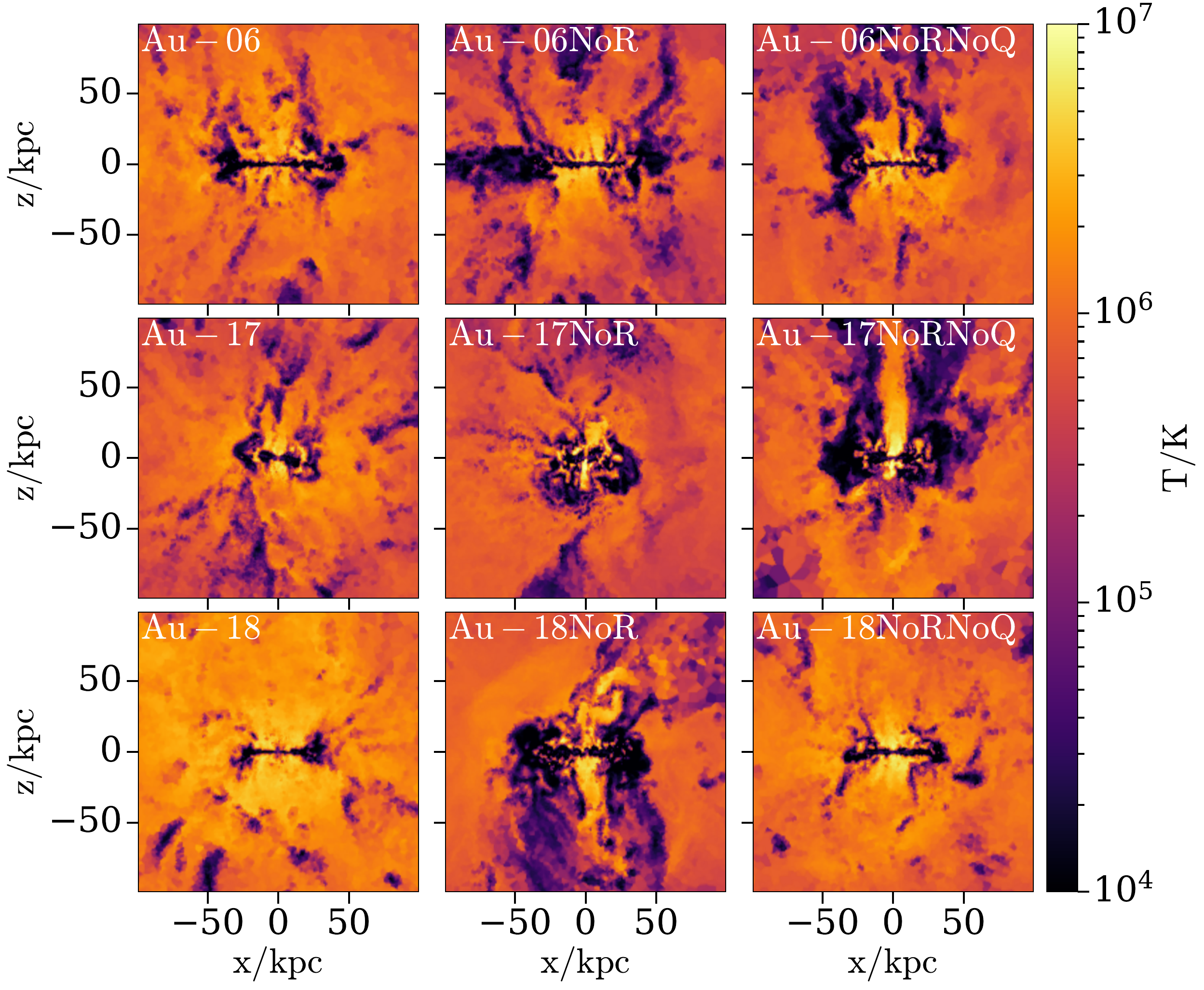}}}$
$\vcenter{\hbox{\includegraphics[width=0.37\textwidth]{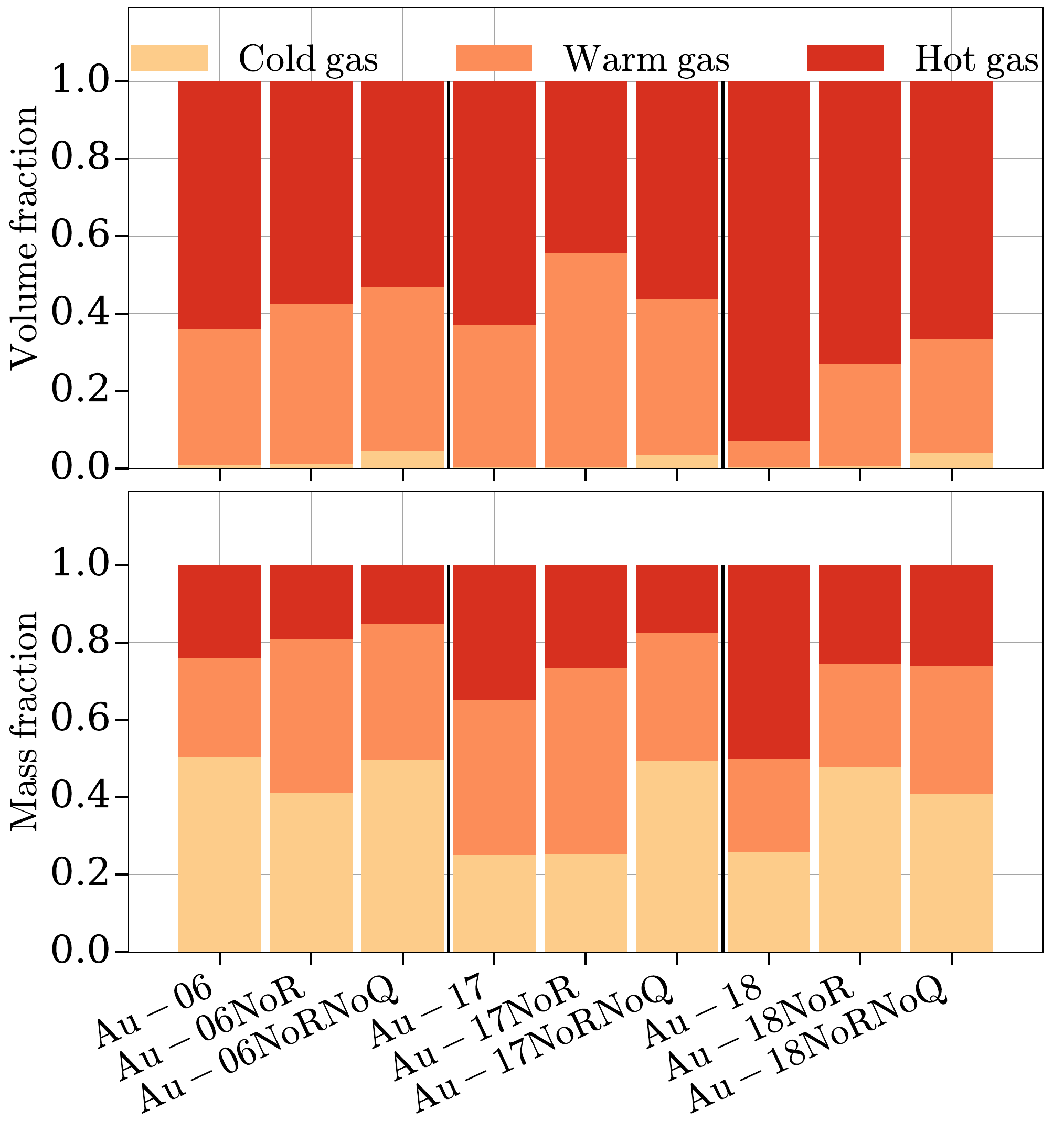}}}$
\caption{\textit{Left-hand plot:} Edge-on projection of the gas temperature for all gas cells inside a 200$\times$200$\times$4 kpc$^3$ box at $z$ = 0. The top, middle, and bottom rows contain results for Au-06, Au-17, and Au-18, respectively. The left-hand, middle, and right-hand columns contain results for the fiducial halo, the NoR, and the NoRNoQ variant, respectively. \textit{Right-hand plot:} Volume-weighted (top row) and mass-weighted (bottom row) fractional breakdown of gas into different temperature regimes inside $R_{200}$. In each stacked bar, the yellow, orange, and red bars represent the fractions of cold ($T$ < 2 $\times$ 10$^4$ K), warm (2 $\times$ 10$^4$ K < $T$ < 5 $\times$ 10$^5$ K), and hot ($T$ > 5 $\times$ 10$^5$ K) gas, respectively. The height of each bar represents the 1 Gyr time-averaged value of each ratio. The lack of AGN feedback reduces the hot gas mass and volume ratios.}
\label{fig:gas_temperature_regimes_combination}
\end{figure*}

\Fig{gas_temperature_regimes_combination} shows on the left-hand plot the edge-on projection of the gas temperature for all gas cells inside a 200$\times$200$\times$4 kpc$^3$ box at $z$ = 0. The top, middle, and bottom rows contain results for Au-06, Au-17, and Au-18, respectively. The left-hand, middle, and right-hand columns contain results for the fiducial halo, the NoR, and the NoRNoQ variant, respectively. The right-hand plot shows the volume-weighted (top row) and mass-weighted (bottom row) fractional breakdown of gas into different temperature regimes inside $R_{200}$. In each stacked bar, the yellow, orange, and red bars represent the fractions of cold ($T$ < 2 $\times$ 10$^4$ K), warm (2 $\times$ 10$^4$ K < $T$ < 5 $\times$ 10$^5$ K), and hot ($T$ > 5 $\times$ 10$^5$ K) gas, respectively. The height of each bar represents the 1 Gyr time-averaged value of each ratio. 

By visually inspecting the left-hand plot we notice that as we move from the fiducial runs to the NoR and NoRNoQ variants the haloes appear to have lower temperatures due to the lack of radio mode AGN feedback. In addition, we see that when we remove the radio mode feedback (i.e. in the NoR variants), the haloes have lower hot gas mass ratios than the fiducial haloes. When we also remove the quasar mode feedback (i.e. in the NoRNoQ variants) the hot gas fraction is further reduced. Furthermore, the NoRNoQ variants have the highest cold gas volume ratios but not necessarily the highest cold gas mass fractions (see below). These features reflect the effect of the radio mode feedback on the gas and indicate that the lack of AGN produces cooler haloes. As discussed in \Sec{Present day galactic properties:Stellar mass distribution}, this results in more extended stellar discs since as shown in \Fig{delta_sfr_regimes_combination} the gas in the outer disc ($R$/kpc > 1) is able to cool quicker and form stars more efficiently. Lastly, the unexpected decrease in the cold gas mass fractions in Au-06NoR and the NoRNoQ variants of Au-06 and Au-18 can be explained by the enhanced star formation rates reported in \Fig{delta_sfr_regimes_combination}, which result in more frequent star formation feedback that prevents the gas from cooling. 

\begin{figure*}
\centering \includegraphics[width=\textwidth]{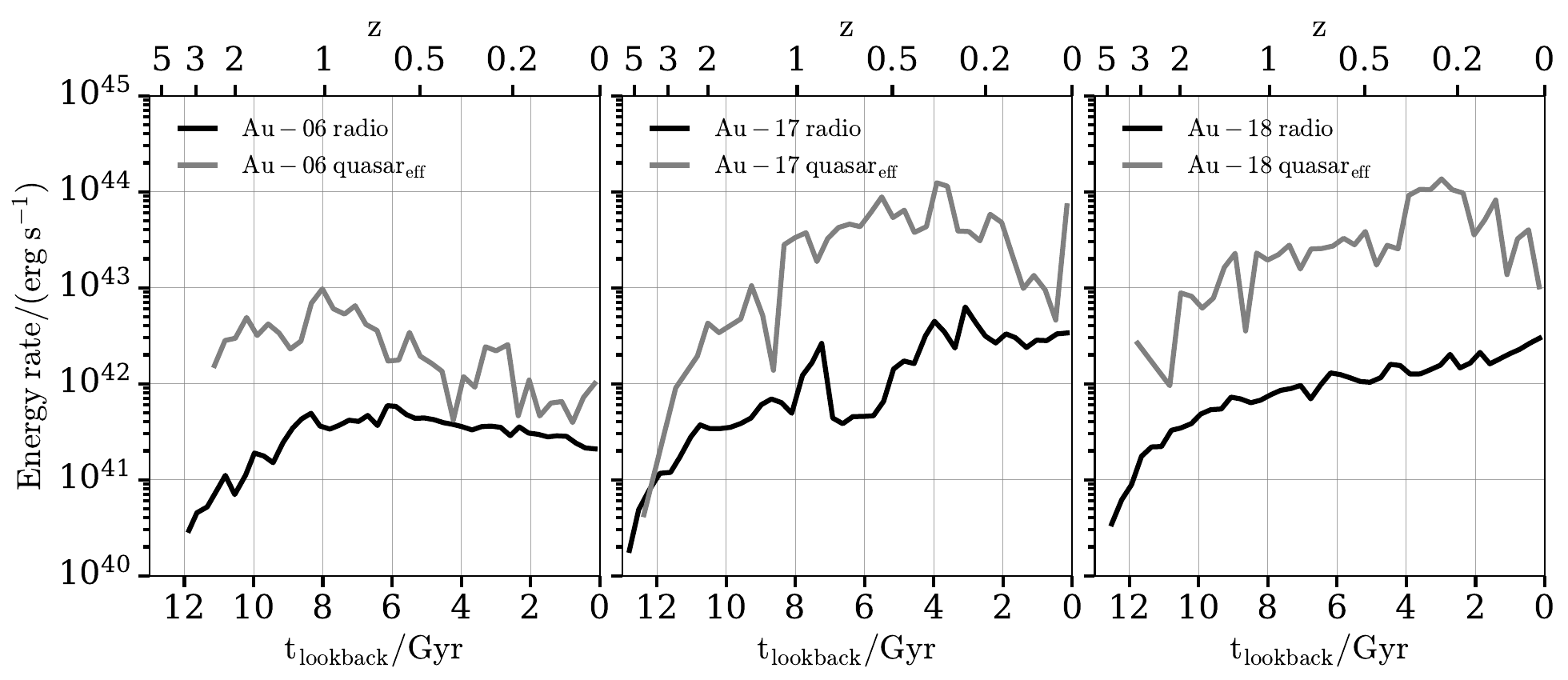}
\caption{Evolution of the radio mode feedback energy rate. The left-hand, middle, and right-hand panels contain results for Au-06, Au-17, and Au-18, respectively. The black curves show the radio mode energy as calculated by the model (see \Sec{The Auriga simulations:Black hole and AGN feedback model}) and the grey curves show the effective quasar mode energy which corresponds to the black lines shown in \Fig{quasar_mode_distribution_combination}. The radio mode feedback is at all times weaker than the quasar mode feedback.}
\label{fig:radio_mode_distribution_combination}
\end{figure*}

\Fig{radio_mode_distribution_combination} shows the evolution of the radio mode feedback energy rate. The left-hand, middle, and right-hand panels contain results for Au-06, Au-17, and Au-18, respectively. The black curves show the radio mode energy as calculated by the model (see \Sec{The Auriga simulations:Black hole and AGN feedback model}) and the grey curves show the effective quasar mode energy as presented in \Fig{quasar_mode_distribution_combination}.

We see that the evolution of the radio mode feedback energy rate shows similar trends with the evolution of the quasar mode shown in \Fig{quasar_mode_distribution_combination}: Au-06 shows a slightly decreasing trend below $z \sim$ 1, whilst Au-17's and Au-18's radio mode feedback energy rates increase as we go to lower redshifts. However, the radio mode is -- by roughly an order of magnitude -- less energetic than the quasar mode. 

In summary, as discussed in \Sec{Quasar mode effects:Star formation rate histories} and shown in \Fig{gas_temperature_regimes_combination} and \Fig{radio_mode_distribution_combination}, the radio mode feedback is less powerful and more responsible for heating up the halo whilst the quasar mode has higher impact and mainly affects the inner regions gas of MW-mass galaxies.

\section{Discussion} \label{sec:Discussion}

In this section we briefly discuss our findings and put them in the context of previous studies.

In \Sec{Present day galactic properties}, we showed that the AGN feedback in the \auriga\ simulations has a significant effect on the structural and dynamical properties of MW-mass galaxies. In particular, including the quasar and radio mode AGN feedback leads to weaker and longer bars, that form later, or do not form at all (see \Fig{bar_strength_profile_combination}). This finding motivated us to explicitly explore in \Sec{Quasar mode effects} and \Sec{Radio mode effects} how the quasar (i.e. ejective) and radio (i.e. preventive) mode feedback, respectively, affect the formation and properties of the bar, bulge, and disc. 

As presented in \Fig{delta_sfr_regimes_combination} and also shown in \cite{GGM17}, the preventive mode does not have a significant effect on the inner galaxy, but rather suppresses star formation in the outer disc. On the other hand, the ejective mode feedback acts to significantly suppress star formation in the central region of the galaxy. Therefore, turning off this mode of feedback leads to galaxies being more baryon-dominated in the inner regions. However, upon carrying out a 2D bar/bulge/disc decomposition we found that this additional mass is in a disc-like rather than a spheroidal component, which results in a more bar-unstable disc in which bars form earlier (see \Fig{bar_strength_profile_combination}).

\subsection{The effects of ejective feedback on the dynamical stability of galaxies} \label{sec:Discussion:The effects of ejective feedback on the dynamical stability of galaxies}

A previous study that explored the bar-AGN connection \citep{BMK16} found that the addition of AGN feedback in the \eris\ simulation (i.e. \erisBH) promotes the development of a bar, since it suppresses the formation of a massive spheroidal bulge at the centre. However, the conclusions we drew in \Sec{Present day galactic properties:Stellar mass distribution} on the basis of \Fig{stellar_surface_density_profiles_combination}, \Tab{stellar_surface_density_profiles_combination_table}, and \Tab{Imfit} indicate that in the \auriga\ simulations the removal of AGN feedback results in low S\'ersic index components at the centre, which do not represent a classical bulge-like population (see also \App{Stellar surface density profiles}). As shown in \Fig{gas_temperature_vs_distance_combination_005}, the lack of AGN feedback allows for gas cells -- which appear to be able to maintain some angular momentum -- to survive in the central regions and settle in a disky component. 

Given that the dominant feedback mode in the central region is the quasar mode (see \Sec{The Auriga simulations} for a description) and that this mode has been implemented in a similar way \citep{DSH05,SDH05} both in the \auriga\ and the \erisBH\ simulations, one would expect to find similar effects of AGN feedback on the formation and properties of bars. However, as described above, we find different results in the two studies, which may be due to a number of different factors. This could include aspects such as different merger histories of the galaxies under study (e.g. a more violent merger history for the \erisBH\ galaxy), the different hydrodynamical schemes employed in the simulations \citep{SWP12,RKH21} or more complex aspects of the physics (sub-grid) modelling employed in the simulation, such as the star formation and feedback \citep[e.g.][discussed the importance of the feedback employed in the simulations on the dynamical properties of barred galaxies]{ZCD19}. Further work would be needed to identify the exact reasons.

\subsection{A complex feedback cycle involving bars and black holes} \label{sec:Discussion:A complex feedback cycle involving bars and black holes}

As briefly discussed in \Sec{Quasar mode effects:Star formation rate histories}, non-linear interactions between bar-driven gas flows, star formation, and AGN feedback occur in barred galaxies. The non-axisymmetric potential generated by a bar exerts torques on the gas \citep{SFB89}  which flows towards the centre, intensifies central star formation (see \Fig{delta_sfr_regimes_combination}), and also accretes onto the black hole. The latter phenomenon will trigger AGN feedback which will heat up and expel the neighbouring gas cells (see \Fig{gas_temperature_vs_distance_combination_005}) and, according to our findings, would subsequently weaken the bar (see \Fig{bar_strength_profile_combination}). Therefore, it becomes apparent that there is a non-trivial feedback cycle which involves bar properties, gas inflows, and AGN activity: as the bar pushes gas to the centre it triggers AGN feedback which could result in weaker bars and consequently less bar-driven gas inflows. Hence, the impact of AGN feedback on the bar-related gas flows via the formation of a weaker bar can be considered as an additional, indirect AGN feedback mechanism which we intent to study in more detail in a future work.

Furthermore, whilst in this work we showed that galaxies with and without radio mode feedback form their bars at different times (i.e. AGN feedback affects the dynamical properties of MW-like galaxies), \Fig{stellar_blackhole_masses_combination} revealed that different variants end up in similar regions on the black hole mass - stellar mass plane at $z$ = 0. This is an important result if we consider that the black hole mass - stellar mass relation is frequently used when calibrating the free parameters of sub-grid models -- especially the efficiency of the black hole accretion/feedback model -- in large scale cosmological simulations \citep[e.g.][]{SCB15,VDP16,PSN18,DAN19,HSG21}. Therefore, we argue that besides fundamental properties like the size, mass, and star formation rate of galaxies; structural parameters, such as the formation time and properties of bars, contain valuable information which can provide additional constraints when calibrating the efficiency of AGN feedback models, since the morphology of MW-mass galaxies can be highly affected by the choice of parameters and modes of AGN feedback (i.e. ejective or preventive).

\subsection{The fate of unbarred galaxies that lack AGN feedback} \label{sec:Discussion:The fate of unbarred galaxies}

This work mainly focused on the effects of AGN feedback on two strongly (Au-17 and Au-18) and one very weakly barred galaxy (Au-06), which can be considered an unbarred galaxy by some bar-strength classification methods. For Au-06 we showed that when we removed its AGN feedback a strong, well-developed bar formed and we could clearly call Au-06NoRNoQ a strongly barred galaxy (see \Fig{bar_strength_profile_combination}). 

A few questions that naturally arise are: Was Au-06 a special case? Would all very weakly barred/unbarred galaxies behave the same? If not, what are the factors that will dictate the fate of unbarred galaxies that lack AGN feedback? Our study showed that when we remove the AGN feedback our galaxies become more baryon-dominated and consequently more bar-unstable, and indeed as Au-06 demonstrated, a weakly barred/unbarred galaxy can become strongly barred. However, whether or not an unbarred galaxy that lacks AGN feedback will form a bar strongly depends on how close that galaxy is to bar instability to begin with, which we cannot reliably assess in advance \citep[e.g.][]{A08}. Thus, in order to be able to generalise our results to all unbarred galaxies we need to investigate in a systematic way the effects of AGN feedback on unbarred galaxies. Fully exploring this parameter space is beyond the scope of this paper but we intend to further investigate the effects of AGN feedback on all \auriga\ galaxies in a future study.

\section{Conclusions} \label{sec:Conclusions}

In this work, we study the impact of the radio (i.e. preventive) and quasar (i.e. ejective) AGN feedback modes on the structural and dynamical properties of barred galaxies. More explicitly, we assess to what extent the \auriga\ AGN implementation affects the properties of the bar/bulge/disc, and consequently of the galaxy as a whole. For that purpose, we selected three \auriga\ galaxies -- two strongly and one very weakly barred -- and re-simulated each one with two different AGN feedback setups: one with no radio mode feedback (NoR) and one with neither the radio mode nor the quasar mode (NoRNoQ). Our main conclusions are as follows:

\begin{itemize}
\setlength{\itemindent}{0.5em}
\item Different stellar patterns (e.g. bars and boxy/peanut bulges) emerge in the central regions of galaxies with different AGN feedback modes (see \Fig{stellar_density_combination}). Hence, we find that the AGN feedback in \auriga\ simulations is able to shape the morphology and affect the structural properties of MW-mass galaxies. This is in contrast to some previous studies which considered AGN feedback to have negligible effect on MW-mass haloes, and thus excluded it from their sub-grid models.

\item While the radio mode feedback does not directly influence the formation and the properties of the bar (it does indirectly by altering the effectiveness of the quasar mode, see \Fig{quasar_mode_distribution_combination}), the quasar mode can have a significant impact on the structural and dynamical properties of barred \auriga\ galaxies. We find that the NoRNoQ variants develop stronger bars (i.e. $A_2$ > 0.3) which form earlier than the bars of the fiducial haloes and the NoR variants (see \Fig{bar_strength_profile_combination}).

\item For all galaxies the stellar masses increase as we move from the fiducial haloes to the NoR and NoRNoQ variants (see \Fig{stellar_vs_halo_mass_combination}). This results from the enhanced star formation rates -- especially in the centre -- due to the reduced AGN feedback (see \Fig{delta_sfr_regimes_combination}).

\item While there is an increase of mass concentration in the centre of the NoR and NoRNoQ variants (see \Fig{stellar_surface_density_profiles_combination}), this is not found to be in a high S\'ersic classical-bulge like component. This result suggests that gas cells which survive and/or accrete in the central regions (due to the lack of the AGN feedback) are able to maintain some angular momentum and settle in a disc component.

\item We find that the quasar and radio mode feedback affect galaxies in the \auriga\ simulations in different ways. The quasar mode feedback mostly influences star formation in the central kiloparsec (see \Fig{delta_sfr_regimes_combination}), by heating up and expelling the gas within the central regions of the galaxy (see \Fig{gas_temperature_vs_distance_combination_005}), i.e. within a sphere with radius \rBH. Even though most of the gas within that sphere is on the effective equation of state -- and thus able to radiate away the energy injected by the quasar mode of the AGN -- there is a non-negligible fraction of gas cells which are able to couple to the quasar mode feedback (see \Fig{AGN_feedback_kernel_combination} and \Fig{quasar_mode_distribution_combination}).

\item The radio mode feedback mostly affects the ability of gas in the halo to cool (see \Fig{gas_temperature_regimes_combination}). Therefore, the radio mode has most of its effect on star formation in the outer parts of the galaxy at late times (see \Fig{delta_sfr_regimes_combination}) and in general leads to galaxies with more extended discs (see \Fig{stellar_density_combination} and \Fig{stellar_surface_density_profiles_combination}).
\end{itemize}

We conclude that AGN feedback can regulate the structural and dynamical galactic properties both directly (by suppressing central star formation) and indirectly (by heating up the halo), and differently affects strongly and weakly barred galaxies. Hence, the way AGN feedback is implemented can play a fundamental role in the properties and morphology of MW-mass galaxies. In addition, we propose that observational studies of the morphology of MW-mass galaxies along with the detailed analysis of their bars can be used as method to quantify the efficiency of ejective AGN feedback.

\section*{Acknowledgements}

The authors would like to thank the anonymous referee for their useful comments which contributed to improve this work. DI would like to acknowledge his family members for their financial support and the European Research Council via ERC Consolidator Grant KETJU (no. 818930). In addition, he gratefully acknowledges the hospitality of the Max Planck Institute for Astrophysics in Garching, Germany, during part of this work, and thanks Jessica May Hislop for valuable discussions. RG acknowledges financial support from the Spanish Ministry of Science and Innovation (MICINN) through the Spanish State Research Agency, under the Severo Ochoa Program 2020-2023 (CEX2019-000920-S). FAG acknowledges financial support from FONDECYT Regular 1211370 and from the Max Planck Society through a Partner Group grant. FAG gratefully acknowledges support  by the ANID BASAL project FB210003. FM acknowledges support through the program ``Rita Levi Montalcini'' of the Italian MUR.

This work used the DiRAC@Durham facility managed by the Institute for Computational Cosmology on behalf of the STFC DiRAC HPC Facility (www.dirac.ac.uk). The equipment was funded by BEIS capital funding via STFC capital grants ST/K00042X/1, ST/P002293/1, ST/R002371/1, and ST/S002502/1, Durham University and STFC operations grant ST/R000832/1. DiRAC is part of the National e-Infrastructure. 

We thank the developers of \astropy\ \citep{AC13,A18}, \matplotlib\ \citep{H07}, \numpy\ \citep{VCG11}, and \scipy\ \citep{VGO20}.

Part of this work was carried out during the COVID-19 pandemic and would not have been possible without the tireless efforts of the essential workers, who did not have the safety of working from their homes.

\section*{Data Availability}

The data underlying this article will be shared on reasonable request to the corresponding author.

\bibliographystyle{mnras}
\bibliography{mn-jour,paper}

\appendix
\section{2D decompositions} \label{app:2D decompositions}

\begin{figure}
\centering 
\includegraphics[width=0.49\textwidth]{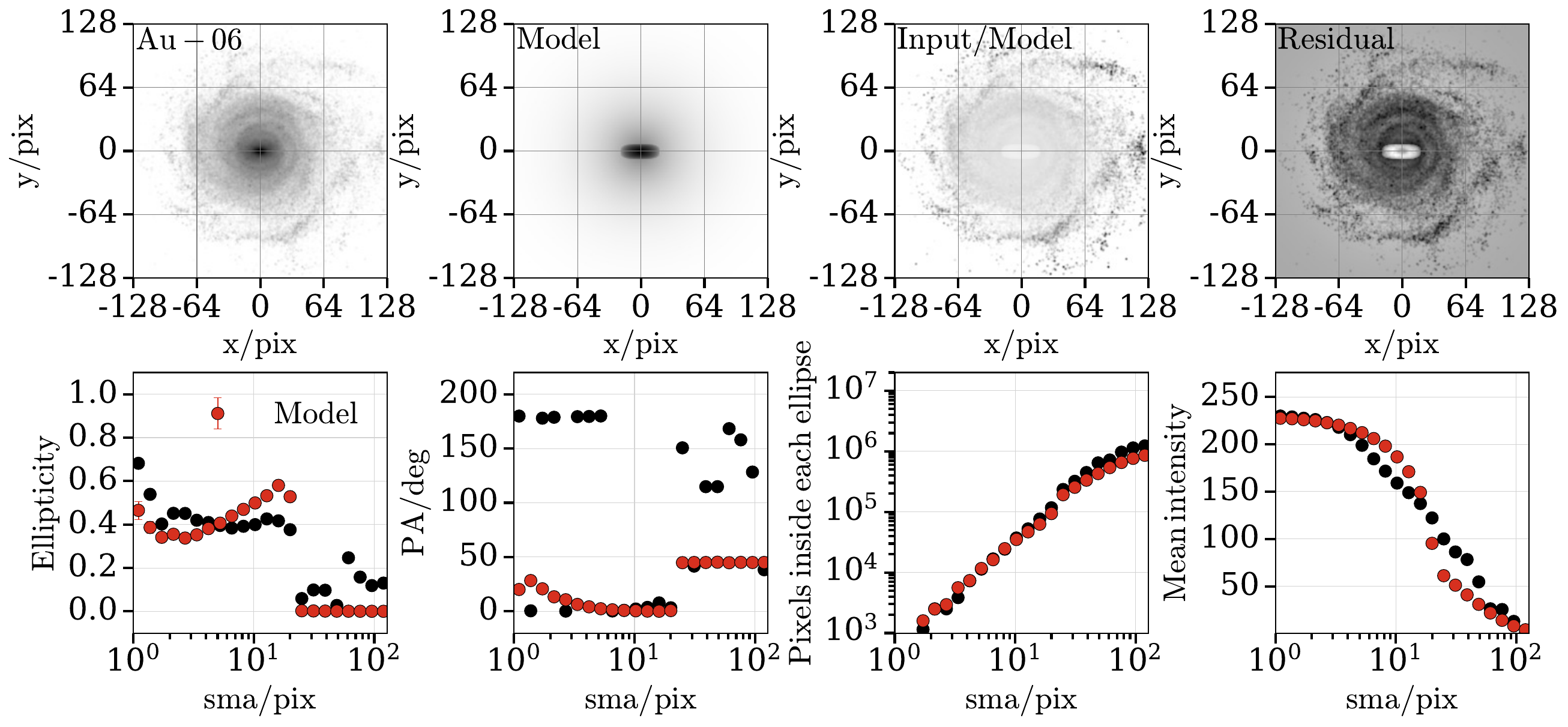}
\includegraphics[width=0.49\textwidth]{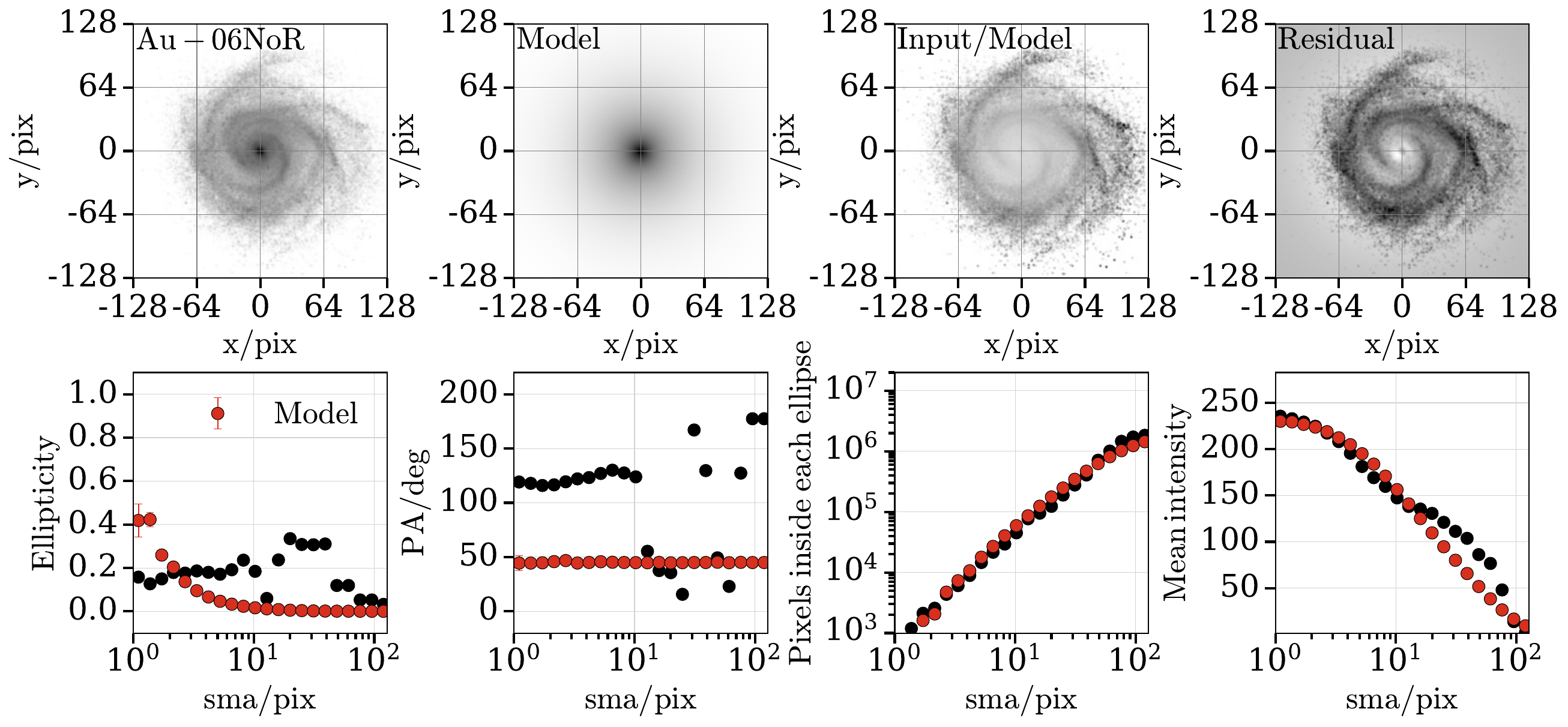}
\includegraphics[width=0.49\textwidth]{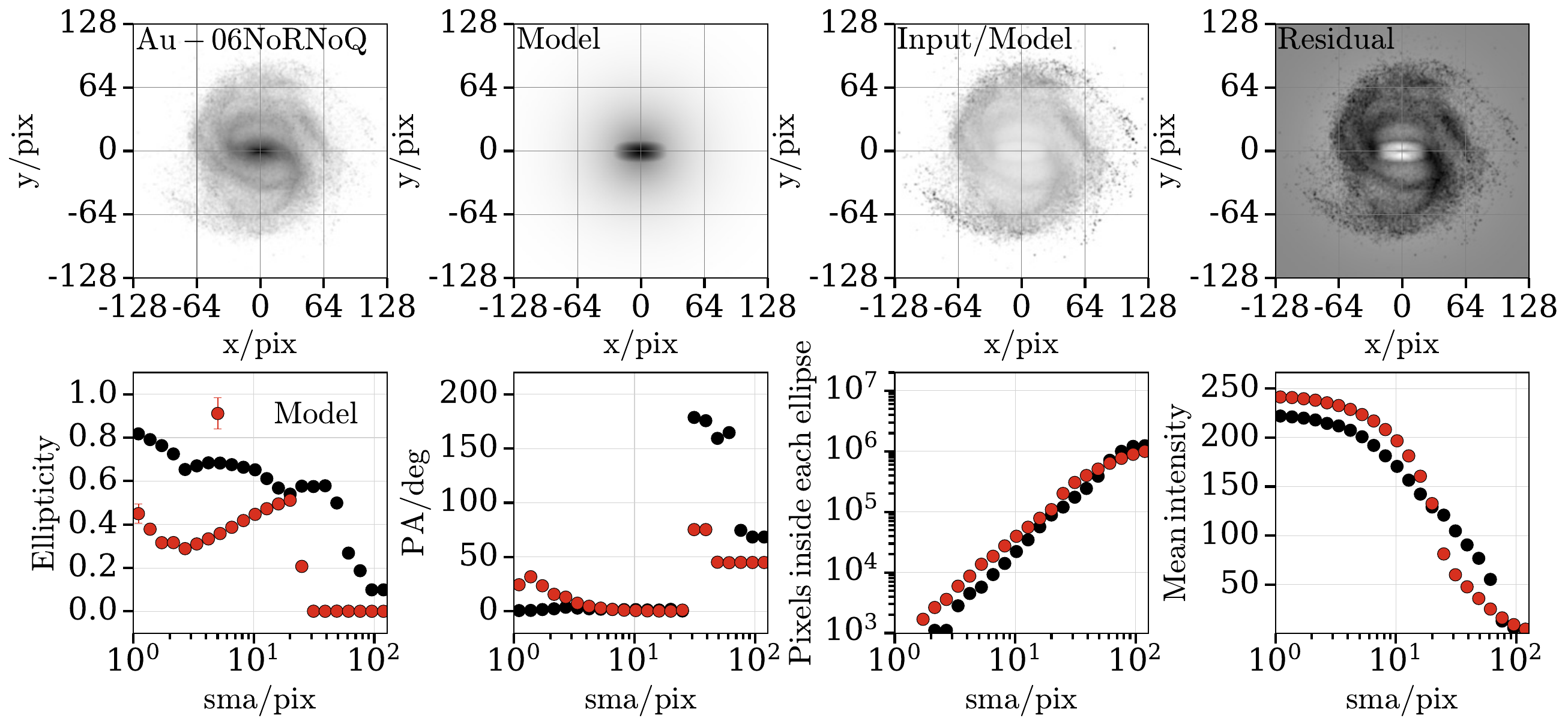}
\caption{2D bar/bulge/disc decomposition for Au-06 and its variants. The top, middle, and bottom plot show the fiducial halo, the NoR, and the NoRNoQ variant, respectively. In each plot the top four panels show from left to right the $r$-band image, model, input/model, and residual produced by \imfit. The bottom four panels show from left to right the ellipticity, position angle, pixel density, and intensity of the $r$-band image (black) and model (red) as a function of the semi-major axis.}
\label{fig:Au-06}
\end{figure}

\begin{figure}
\centering 
\includegraphics[width=0.49\textwidth]{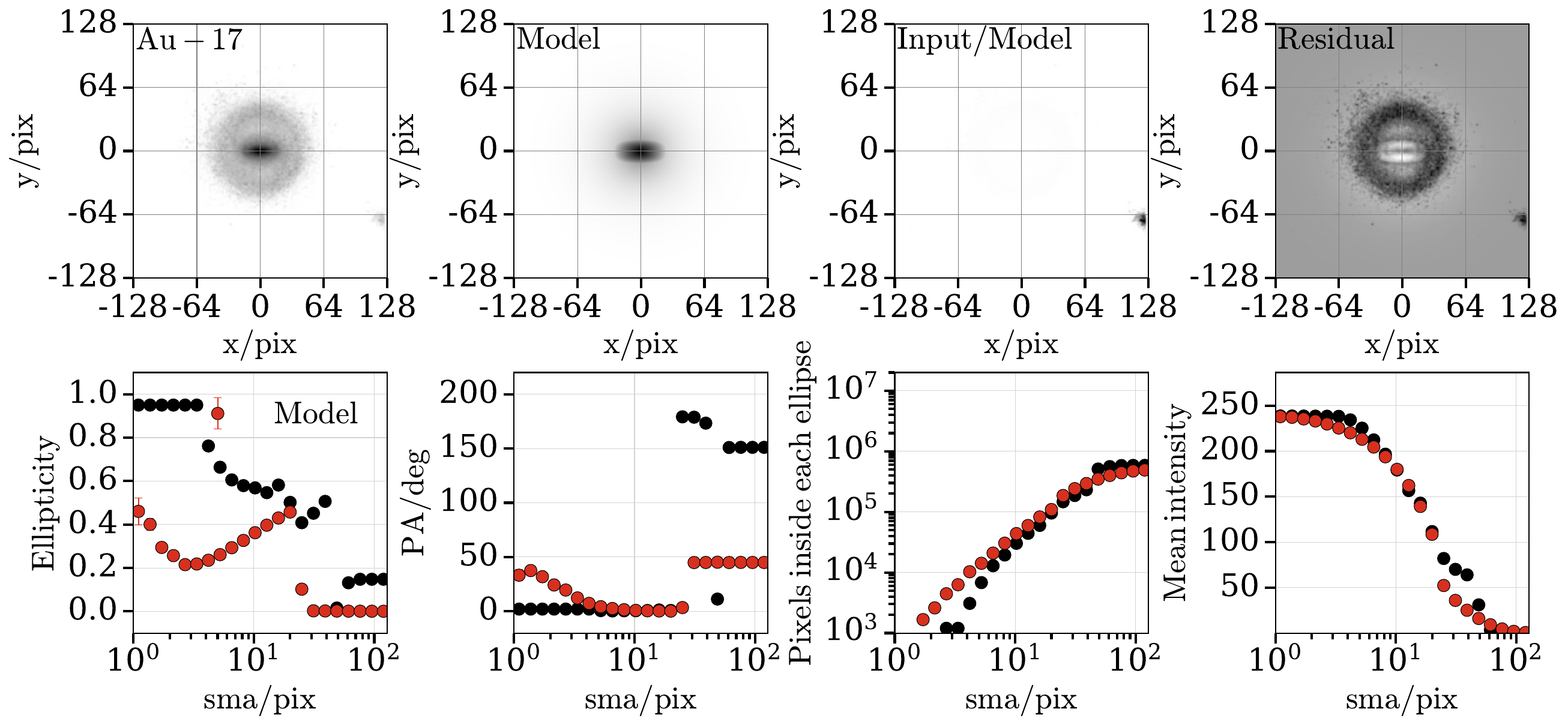}
\includegraphics[width=0.49\textwidth]{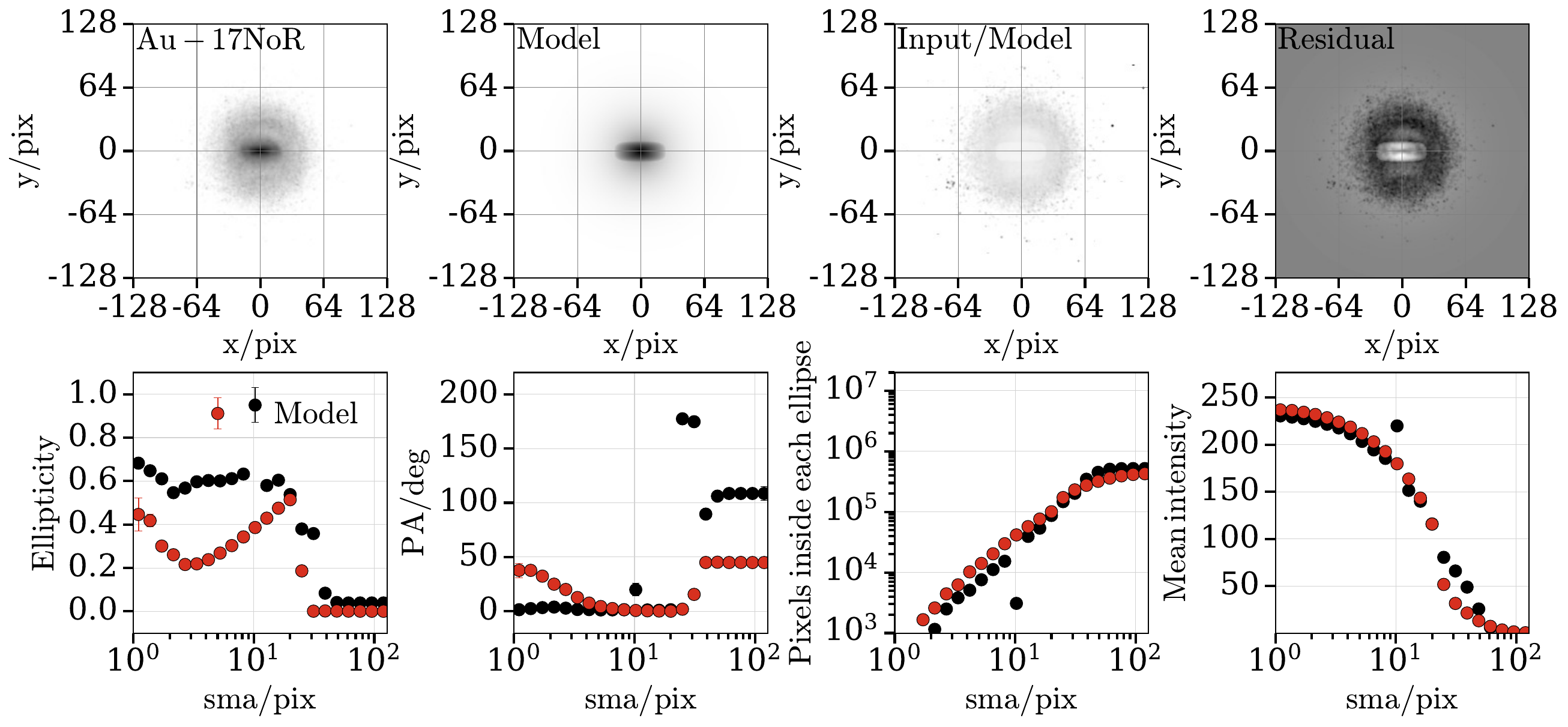}
\includegraphics[width=0.49\textwidth]{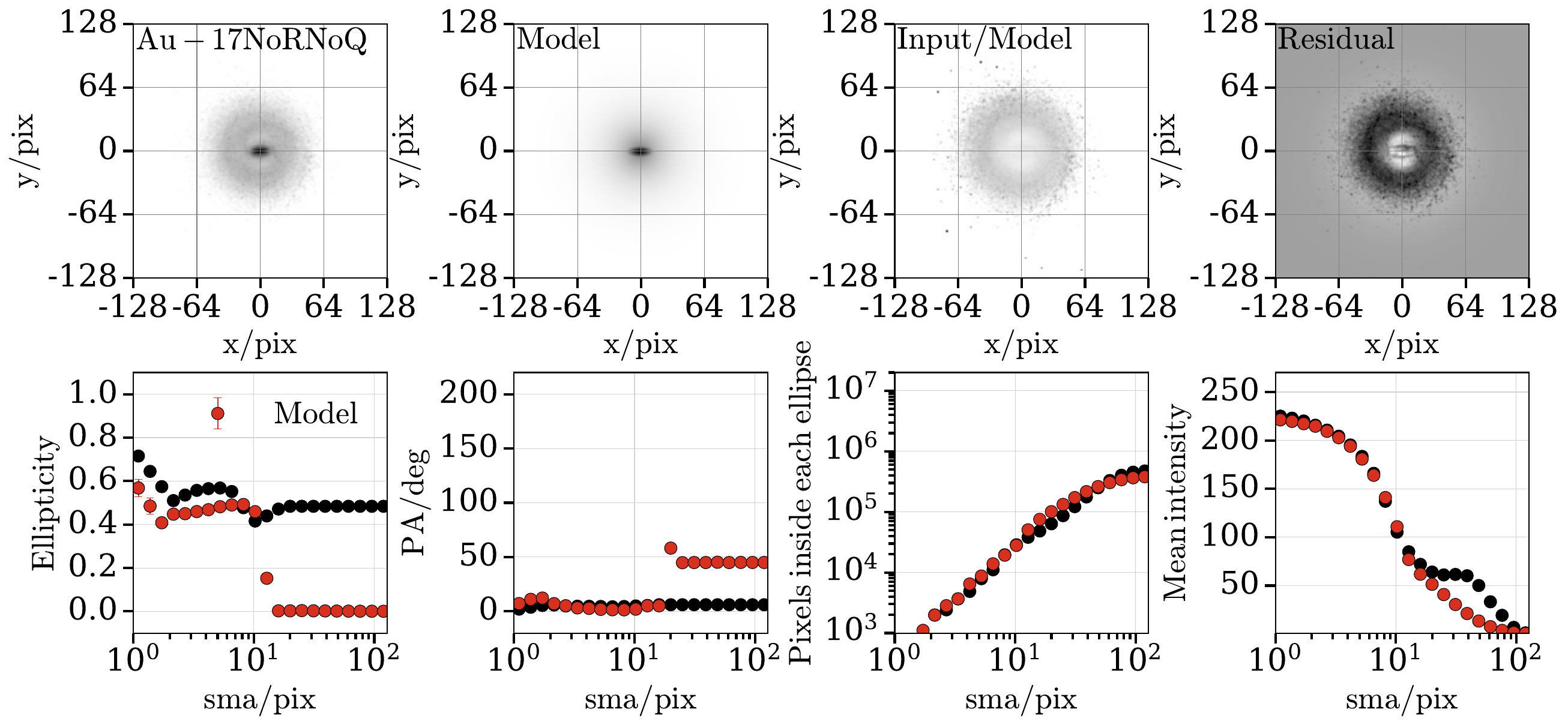}
\caption{Same as \Fig{Au-06} but for Au-17 and its variants.}
\label{fig:Au-17}
\end{figure}

\begin{figure}
\centering 
\includegraphics[width=0.49\textwidth]{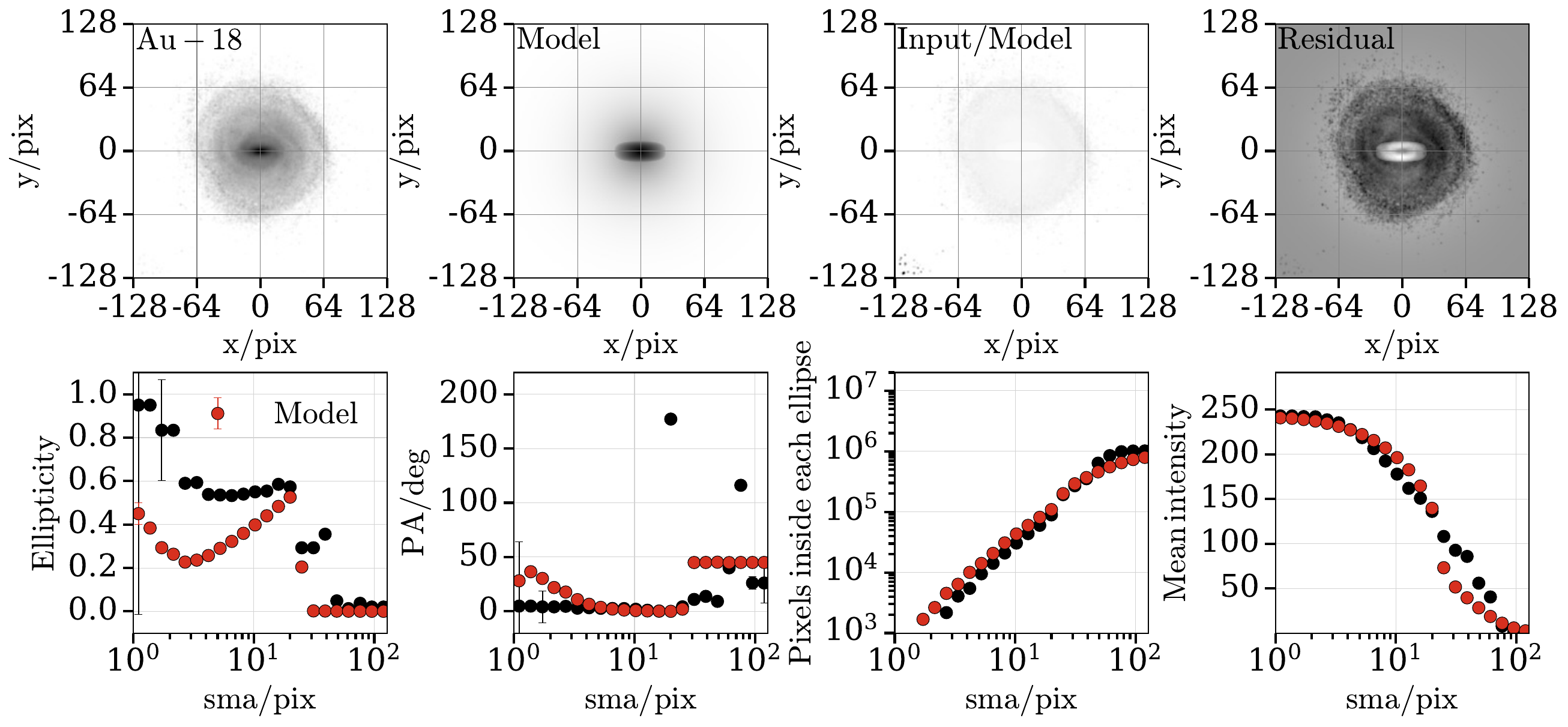}
\includegraphics[width=0.49\textwidth]{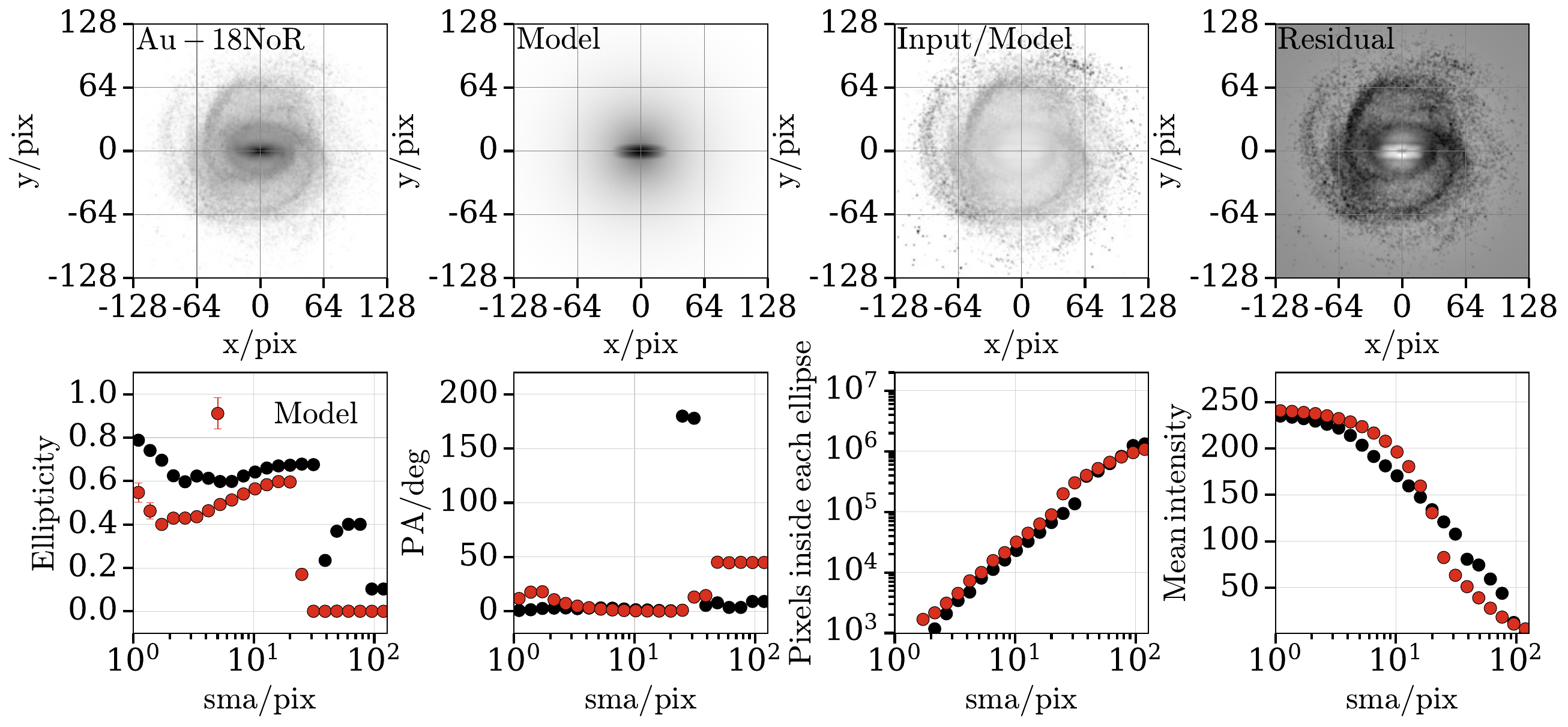}
\includegraphics[width=0.49\textwidth]{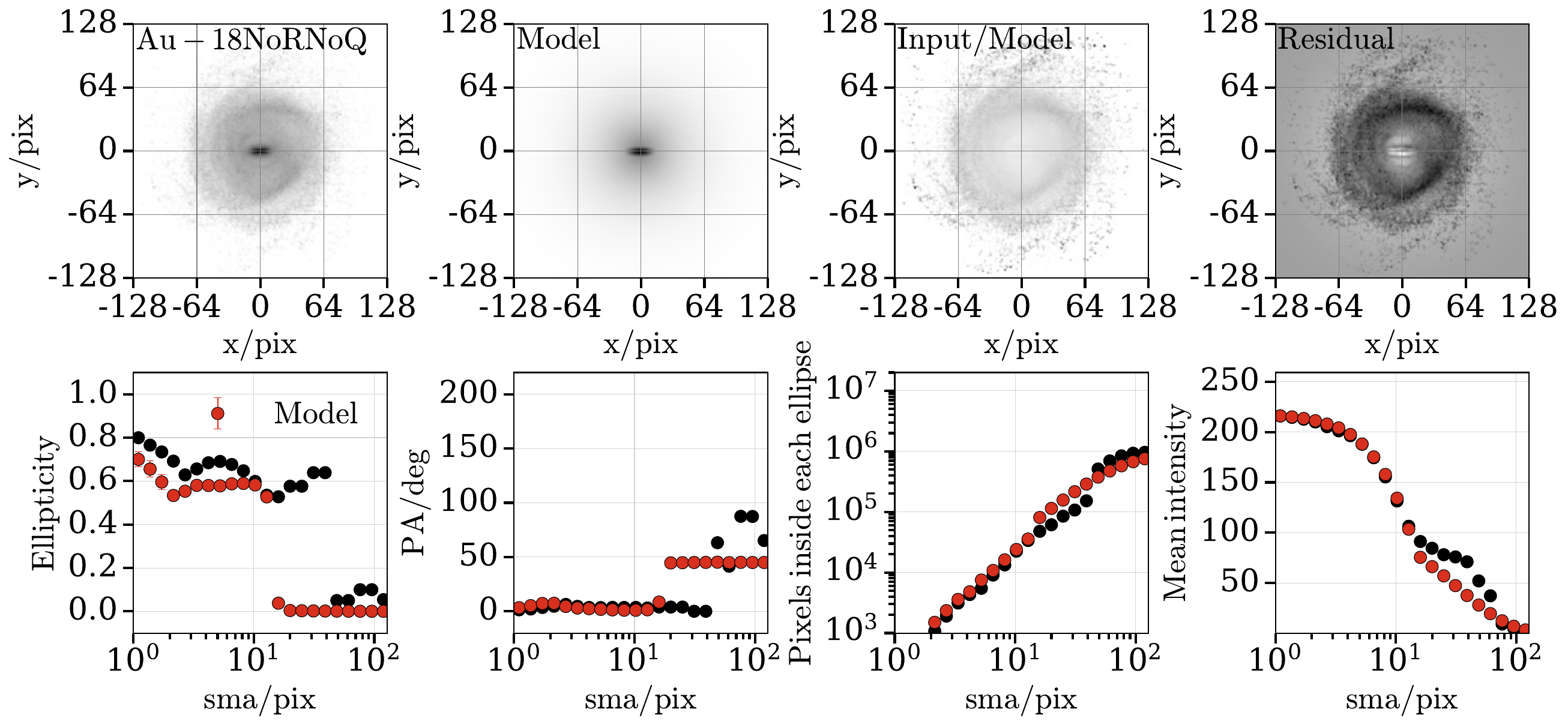}
\caption{Same as \Fig{Au-06} but for Au-18 and its variants.}
\label{fig:Au-18}
\end{figure}

The fitting process begins by producing face-on images for each halo and gaining a first insight into each galaxy's parameters. We do that as follows:
\begin{enumerate}[wide=0pt, label=(\roman*)]
\item We rotate all stellar particles such that the bar is along the horizontal axis.
\item We create face-on grey-scaled $r$-band images.
\item We use a Gaussian filter with FWHM = 2 to blur the images, in order to appropriately simulate the effects of seeing in real observations.
\item We use \photutils\ \citep{BSR19} to perform isophotal ellipse fitting to the images and estimate the intensity, length, and ellipticity of the bar ($I_0, \abar, ell$), bulge ($\Ie, \re, ell$), and disc ($I_0, h, ell$) which we provide to \imfit\ as an initial set of parameters for the corresponding profiles.
\end{enumerate}

\Fig{Au-06}, \Fig{Au-17}, and \Fig{Au-18} contain 2D bar/bulge/disc fits for Au-06, Au-17, and Au-18, respectively. In each figure the top, middle, and bottom plot show the fiducial halo, the NoR, and the NoRNoQ variant, respectively. In each plot the top four panels show from left to right the $r$-band image, model, input/model, and residual produced by \imfit. The bottom four panels show from left to right the ellipticity, position angle, pixel density, and intensity of the $r$-band image (black) and model (red) as a function of the semi-major axis.

\section{Evolution of the gas temperature-distance relation} \label{app:Evolution of the gas temperature-distance relation}

\begin{figure}
\centering 
\includegraphics[width=0.49\textwidth]{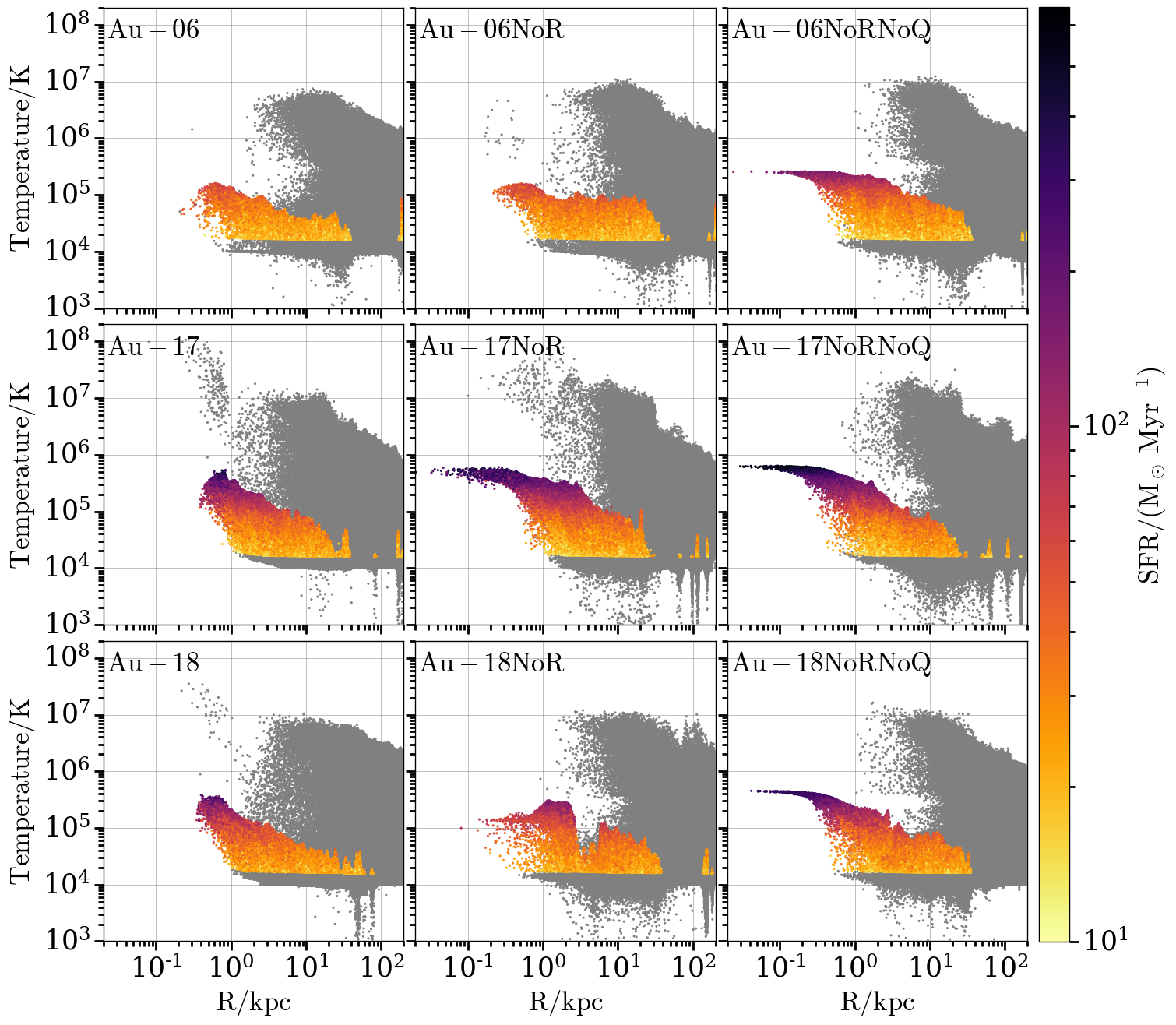}
\caption{Same as \Fig{gas_temperature_vs_distance_combination_005} but at $z$ = 0.00.}
\label{fig:gas_temperature_vs_distance_combination_0}
\end{figure}

\begin{figure}
\centering 
\includegraphics[width=0.49\textwidth]{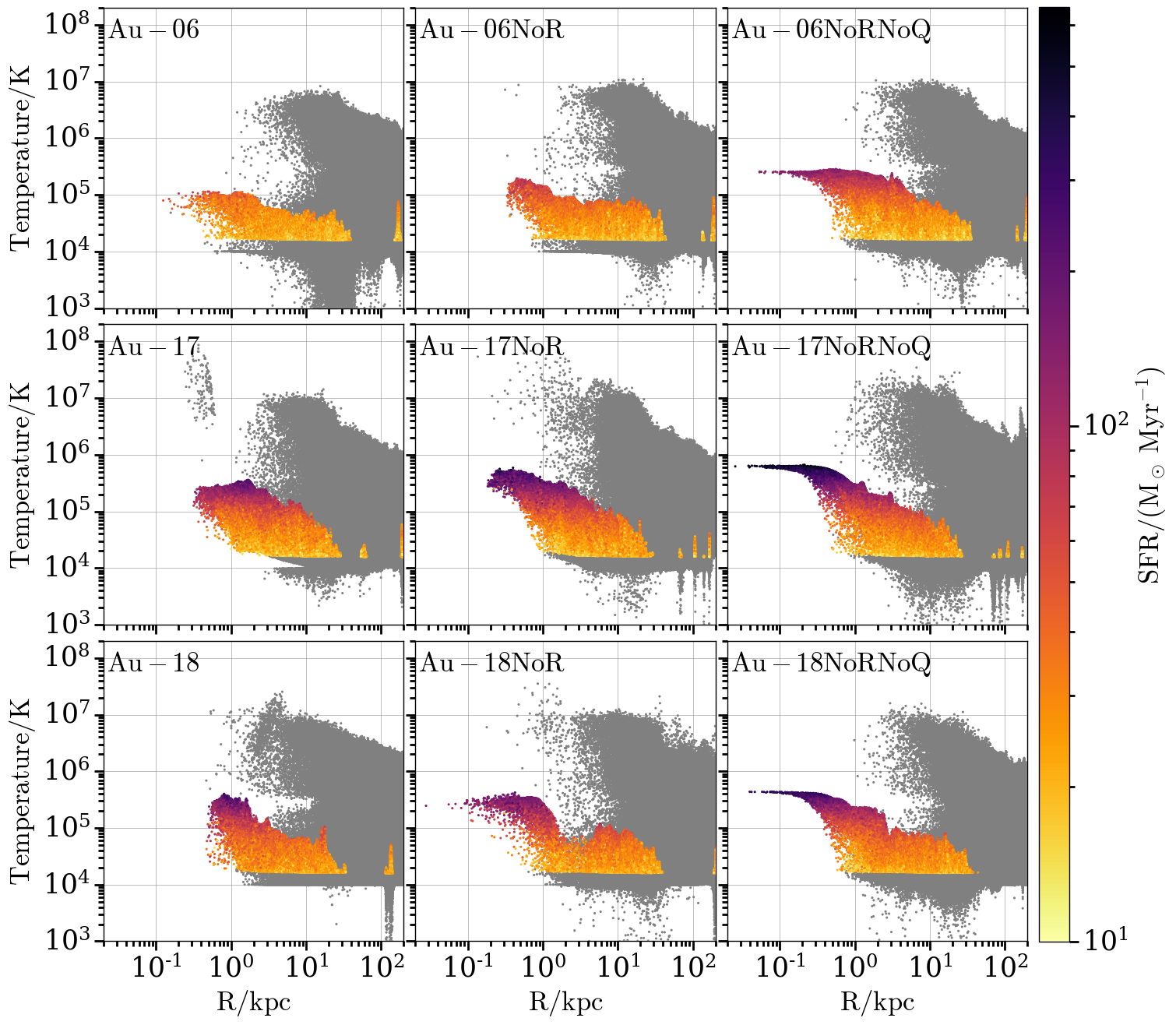}
\caption{Same as \Fig{gas_temperature_vs_distance_combination_005} but at $z$ = 0.02.}
\label{fig:gas_temperature_vs_distance_combination_002}
\end{figure}

In \Fig{gas_temperature_vs_distance_combination_0} and \Fig{gas_temperature_vs_distance_combination_002} we reproduce \Fig{gas_temperature_vs_distance_combination_005} for $z$ = 0.00 and $z$ = 0.02, respectively, to show that even though in all three plots the horn-like features discussed in \Sec{Quasar mode effects:Gas properties} consistently appear in all NoRNoQ variants; at $z$ = 0.00 Au-17NoR also shows the same feature (i.e. has gas cells on the eEoS). This behaviour shows that despite the fact that its quasar mode is on, it is not ejecting energy in this specific snapshot so star forming gas cells exist in the centre as they do in all NoRNoQ variants.

\section{Stellar surface density profiles} \label{app:Stellar surface density profiles}

\begin{table*}
	\centering
	\caption{Similar to \Tab{stellar_surface_density_profiles_combination_table} but instead of $z$ = 0 the fitting was performed at the time the NoRNoQ variant of each halo formed a bar. The rows represent 1) model names (three fiducial runs, namely Au-06, Au-17, and Au-18, plus a no-radio and a no-radio-no-quasar variant for each one); 2) lookback time at which the decomposition was performed; 4) S\'ersic index; 5) effective radius; 6) inferred S\'ersic mass; 8) scale length; 9) inferred disc mass; 10) disc-to-total stellar mass ratio.}
	\label{tab:stellar_surface_density_profiles_combination_table_2}
	\begin{tabular}{lccccccccc} 
		\hline
		  & Au-06 & Au-06NoR & Au-06NoRNoQ & Au-17 & Au-17NoR & Au-17NoRNoQ & Au-18 & Au-18NoR & Au-18NoRNoQ \\
		\hline
		\tlookback/ Gyr & 3.98 & 3.98 & 3.98 & 10.46 & 10.46 & 10.46 & 8.89 & 8.89 & 8.89 \\
		\textbf{S\'ersic} & & & & & & & & & \\
        $n$ & 0.78 & 0.85 & 0.91 & 1.49 & 0.40 & 0.39 & 0.10 & 0.54 & 0.33 \\
        \Reff/kpc & 1.02 & 0.66 & 0.72 & 2.16 & 0.41 & 0.38 & 0.73 & 0.55 & 0.36 \\
        \MSersic/(10$^{10}$ \Msun) & 0.50 & 0.42 & 0.84 & 0.05 & 0.02 & 0.09 & 0.03 & 0.15 & 0.24 \\
		\textbf{Exponential} & & & & & & & & & \\
	    $h$/kpc& 3.84 & 3.96 & 3.06 & 1.73 & 2.01 & 1.86 & 1.91 & 2.29 & 1.92 \\
        \Md/(10$^{10}$ \Msun) & 3.72 & 4.34 & 5.92 & 1.49 & 1.56 & 1.70 & 2.56 & 2.23 & 3.21 \\
        $D/T$ & 0.88 & 0.91 & 0.88 & 0.97 & 0.99 & 0.95 & 0.99 & 0.94 & 0.93 \\
		\hline
	\end{tabular}
\end{table*}

\Tab{stellar_surface_density_profiles_combination_table_2} shows the fitting parameters of the process described in \Sec{Present day galactic properties:Stellar mass distribution} but instead of $z$ = 0 (see \Tab{stellar_surface_density_profiles_combination_table}), here we decompose every version of a halo (i.e. the fiducial halo, the NoR, and the NoRNoQ variant) at the earliest time one of the versions formed a bar. Since as we showed in \Fig{bar_strength_profile_combination} the NoRNoQ variants form bars earlier than their fiducial haloes and NoR variants, in practice we decompose every version of a particular halo at the time the NoRNoQ variant of that halo formed a bar. This allows us to investigate (i) why do the NoRNoQ variants form bars earlier than their fiducial haloes and the NoR variants, and (ii) what are the structural differences between the NoRNoQ variants and their fiducial haloes and NoR variants at the time the NoRNoQ variants form bars.

In \Tab{stellar_surface_density_profiles_combination_table_2}, we can see that the NoRNoQ variants have the most massive S\'ersic and exponential components (as expected since these variants have the highest total stellar masses) and at the same time the lowest $D/T$ ratios\footnote{The $D/T$ values for Au-06 and Au-06NoRNoQ both appear as 0.88 but the actual values are 0.8811 and 0.8793, respectively. Thus, Au-06NoRNoQ has slightly lower $D/T$ hence slightly higher $B/T$ ratio.} (i.e. highest $B/T$ ratios). In addition, if we only take into account haloes that are barred at $z$ = 0 (i.e. all haloes except for Au-06NoR), we see that the effective radii of the NoRNoQ variants are the smallest. Therefore, these central components apart from more massive are also more centrally concentrated whilst maintaining low S\'ersic indices. These results agree with what we reported in \Sec{Present day galactic properties:Stellar mass distribution} and discussed in \Sec{Discussion}, and indicate that these central components are more disc-like components which promote bar instabilities, rather than high-S\'ersic index, dispersion-dominated bulges which would have suppressed the formation of the bar.

\label{lastpage}
\end{document}